\documentstyle[graphics,psfig]{mn}

\def\kms{{\rm km s}^{-1}}
\def\pcm{P$^3$M}\def\pa{\partial}
\def\ap3m{AP$^3$M}
\def\pcm{P$^3$M}\def\pa{\partial}
\def\LCDM{$\Lambda$CDM}
\def\hkpc{h^{-1}\,{\rm kpc}}
\def\hMpc{h^{-1}\,{\rm Mpc}}

\def\d{{\rm d}}
\def\p{\partial}
\def\i{\relax\ifmmode{\rm i}\else\char16\fi}
\def\e{\rm e}
\def\lesssim{{_ <\atop{^\sim}}}
\def\lta{\lesssim}

\def\fracj#1#2{{\textstyle{#1\over#2}}}
\def\b#1{{\bf{#1}}}
\def\ea{et~al.~}                            

\def\lesssim{\mathrel{\hbox{\rlap{\hbox{\lower4pt\hbox{$\sim$}}}\hbox{$<$}}}}
\def\gtrsim{\mathrel{\hbox{\rlap{\hbox{\lower4pt\hbox{$\sim$}}}\hbox{$>$}}}}

\newcommand{\AAA}[3]    {\mbox{#3, A\&A~\textbf{#1},~#2}}

\newcommand{\ApJ}[3]    {\mbox{#3, ApJ~\textbf{#1},~#2}}
\newcommand{\ApJS}[3]   {\mbox{#3, ApJ~Suppl.~\textbf{#1},~#2}}
\newcommand{\ApJL}[3]   {\mbox{#3, ApJ~Lett.~\textbf{#1},~#2}}

\newcommand{\AJ}[3]     {\mbox{#3, Astron.~J.~\textbf{#1},~#2}}
\newcommand{\MNRAS}[3]  {\mbox{#3, MNRAS~\textbf{#1},~#2}}
\newcommand{\Nature}[3] {\mbox{#3, Nature~\textbf{#1},~#2}}

\newcommand{\PhRevL}[3] {\mbox{#3, Phys.~Rev.~Lett.~\textbf{#1},~#2}}

\newcommand{\astroph}[1]{\mbox{\texttt{astro-ph/#1}}}

\begin{document}

   \title[MLAPM a \textit{C} code for cosmological simulations]
	{Multi-Level Adaptive Particle Mesh (MLAPM): \\
          A C code for cosmological simulations}

   \author[Knebe A., Green A. \& Binney J.J.]
          {Alexander Knebe$^1$,
           Andrew Green
           and
           James Binney\\
           Theoretical Physics, 1 Keble Road, Oxford OX1 3NP\\
           $^1$a.knebe1@physics.ox.ac.uk
          }

   \date{Received ...; accepted ...}

   \maketitle

\begin{abstract}
We present a computer code written in \textit{C} that is designed to
simulate structure formation from collisionless matter.  The code is purely
grid-based and uses a recursively refined Cartesian grid to solve Poisson's
equation for the potential, rather than obtaining the potential from a
Green's function.  Refinements can have arbitrary shapes and in practice
closely follow the complex morphology of the density field that evolves. The
timestep shortens by a factor two with each successive refinement.

Competing approaches to $N$-body simulation
are discussed from the point of view of the basic theory of $N$-body
simulation. It is argued that an appropriate choice of softening length
$\epsilon$ is of great importance and that $\epsilon$ should be at all
points an appropriate multiple of the local inter-particle separation.
Unlike tree and \pcm\ codes, multigrid codes automatically satisfy this
requirement. We show that at early times and low densities in cosmological
simulations, $\epsilon$ needs to be significantly smaller relative to the
inter-particle separation than in virialized regions.  Tests of the ability
of the code's Poisson solver to recover the gravitational fields of both
virialized halos and Zel'dovich waves are presented, as are tests of the
code's ability to reproduce analytic solutions for plane-wave evolution. The
times required to conduct a \LCDM\ cosmological simulation for various
configurations are compared with the times required to complete the same
simulation with the ART, A\pcm\ and GADGET codes. The power spectra, halo
mass functions and halo-halo correlation functions of simulations conducted
with different codes are compared.

The code may be down-loaded through one of the authors' web pages.
\end{abstract}


\section{Introduction}
 Over the last two and a half decades great strides have been taken in
understanding the origin of the large-scale structure of the Universe, and
the formation of galaxies. A picture has emerged in which contemporary
structures have evolved by gravitational amplification of seed
inhomogeneities that are likely of quantum origin. This picture ties
together measurements of the cosmic background radiation, estimates of the
primordial abundances of the light elements, measurements of the clustering
of galaxies and, to a more limited extent, the characteristic properties of
individual galaxies.

This picture rests on some important assumptions that have yet to be
convincingly verified. The most important of these is that baryons
contribute only a small fraction of the mean energy density in the Universe,
the bulk being made up of some combination of vacuum energy and dark matter.
Dark matter plays a central role in structure formation because only gravity
couples it to the cosmic background radiation, so it is already free to
cluster in the radiation-dominated era, when baryons are effectively locked
to the relatively incompressible radiation fluid. Consequently, at the era
of decoupling, when the observable baryons are at last able to cluster, they
quickly fall into ready-made structures in the dark-matter density field.

Since dark matter does not interact electromagnetically, it is either
collisionless, or very nearly so (Spergel 2000), and it usually modelled
under the assumption that it is completely collisionless. Consequently, the
governing equations that one needs to solve in order to follow the evolution
of dark matter are the coupled collisionless Boltzmann and Poisson
equations. The standard technique for solving this system is $N$-body
simulation. The purpose of this paper is to present a new code, written in
$C$, for carrying out such simulations in a cosmological context.

Section 2 explains why we think it is important to add another $N$-body code
to the significant numbers of codes that are already available for
cosmological simulations. Section 3 reviews the fundamental principles of
$N$-body simulations in order to clarify the spatial resolution that is
appropriate with a given number of particles. Readers who are already
convinced of the value of our code, and are confident that they understand
what an $N$-body code does, can skip straight to Section 4, which describes
how our multigrid Poisson-solver works. Section 5 describes our algorithm
for advancing particles with multiple timesteps. Section 6 describes and
tests the time-integration scheme employed. Section 7 presents timing data
and energy-conservation data for realistic \LCDM\ simulations. We close with
a discussion of our main results in Section 8.

\section{Why another $N$-body code?}

Since the pioneering simulations in the 70's (e.g., Peebles, 1970;
Haggerty \& Janin, 1974; Press \& Schechter, 1974; White, 1976;
Aarseth, Turner \& Gott, 1979), a great deal of effort has gone into
producing powerful $N$-body codes for cosmological simulations.  The
first simulations evaluated the forces on particles by direct
summation of the Newtonian interaction between particle pairs, but
this is dreadfully inefficient with more than a thousand
particles. Tree codes (Appel 1985; Barnes \& Hut 1986; Dehnen 2000)
radically reduce the cost by grouping distant particles into
aggregates, and then summing over such aggregates rather than over
individual particles.  Particle-Mesh (PM) codes (Hohl 1978; Hockney \&
Eastwood 1988) estimate the density on a grid and then use discrete
Fourier transforms (DFTs) to convolve the density with the Green's
function.  This technique greatly facilitates the imposition of
periodic boundary conditions but suffers from the limitation that the
use of DFTs mandates the use of a regular grid, and such a grid cannot
adequately represent a highly clustered distribution of particles: if
in a low-density region there are a reasonable number of particles in
each cell, high-density regions will be under-resolved; conversely, if
in a high-density region there are a reasonable number of particles in
a cell, in low-density regions nearly all cells will be empty.  Empty
cells are problematic algorithmically (the density is not really zero
at their locations) and represent an unacceptable waste of computer
memory.

In a particle-particle-particle-mesh (\pcm) code, a PM calculation that uses
a coarse grid yields the long-range component of the forces, while direct
summation of additional forces from near neighbours completes the
calculation  (Hockney \& Eastwood 1988; Efstathiou \ea
1985). As clustering develops, large
numbers of particles accumulate in a few cells of a \pcm\ code's coarse
grid, and the direct summation part of the calculation becomes prohibitively
costly. In an adaptive \pcm\ (A\pcm) code this situation is remedied by
replacing the direct summation in a region of high density by an additional
\pcm\ calculation, in which a fine grid covers only the dense region
(Couchman 1991; Couchman \ea 1995). When clustering reaches the point at
which the direct sum of this daughter calculation becomes costly, it is
itself partially replaced by a \pcm\ calculation, and so on indefinitely.

The grid of a \pcm\ code is used only to find the long-range component of
the force. With a sufficiently adaptive grid the entire force can be
calculated on the grid.  Immediately apparent advantages of adaptive grids are that
they naturally admit (i) periodic boundary conditions, (ii) adaptive softening, and
(iii) individual time-steps. Moreover, they provide a framework in which to
do grid-based hydrodynamics.

In view of the potential of adaptive-grid technology, several groups have
tried it for cosmological simulations.  Gnedin (1995) and Pen (1998) start
with a Cartesian grid and let it distort so as to increase resolution in
some regions. This procedure has the drawback of producing significantly
non-cubical cells. Norman \& Bryan (1998) enhance the resolution of a basic
Cartesian grid by placing finer grids over dense regions. These refinements
have to be cubical, and cannot be overlapping. Consequently, large numbers
of small grids would be required to closely follow a highly irregular density
distribution of the type that gravitational clustering generates (cf.\ 
Fig.~\ref{slicefig} below).

We have developed a code, MLAPM, that starts from a regular Cartesian grid
and recursively refines cells such that subgrids can have arbitrary geometry
(subject to each cell being cubical).  MLAPM, which uses a multigrid
algorithm to solve Poisson's equation, is in many ways similar to the
Adaptive~Refinement~Tree (ART) code of Kravtsov et al. (1997, 1999) which also
utilizes recursively placed refinements of arbitrary shape as the simulation
evolves. In Section~\ref{LCDM} we compare the performance of the two codes.
A significant difference between the two codes is that ART, but not MLAPM,
organizes cells into a tree structure -- hence its name.\footnote{However, its
principles are entirely different from those of a conventional Barnes--Hut
tree code.}  We believe that the adaptive multigrid approach is an important
one that should be developed independently by more than one group.

Currently large cosmological $N$-body simulations are being run with
tree, A\pcm\ and multigrid codes. Three considerations will determine which
technology has the biggest impact in the future. One is the importance of
adaptive softening discussed below.  Another is ease of parallelization,
since we are entering an era in which massively parallel computers lie
within the budgets of single research groups. The final consideration is the
ease of including baryons in cosmological simulations. If dark matter exists
and is collisionless, we have a fair idea of how it will cluster. Our
understanding of galaxy formation is, by contrast, very incomplete, and the
future of numerical cosmology lies with simulations that include baryons. 

Our poor understanding of galaxy formation arises in part because
baryons, being dissipative, cluster much more strongly than dark matter, and
galaxies form from the most strongly clustered component. So exquisite
spatial resolution is required to simulate galaxy formation. Several groups
are currently working on ways to include  gas dynamics in cosmological
simulations. [See Frenk et al.\ (1999) for a recent comparison of such codes.] Some use the grid-less approach of
Smooth Particle Hydrodynamics (SPH; Gingold \& Monaghan,
1977; Lucy, 1977), but many use a grid-based scheme. In
developing a grid-based Poisson solver we are in part motivated by the
thought that once the substantial investment required to establish a
dynamical grid has been made, it will be comparatively straightforward
to extend the code to include grid-based hydrodynamics.

\section{Theoretical basis of  $N$-body simulation}

\subsection{Standard $N$-body simulation}

When used to model the dynamics of a collisionless system, an $N$-body code
solves the collisionless Boltzmann equation by the method of characteristics
(e.g., Leeuwin, Combes \& Binney 1993). The characteristics, on which
the phase-space density $f$ is constant, are the possible trajectories of
particles in the system's gravitational potential, $\Phi$. Their integration
requires repeated solution of the Poisson equation
 \begin{equation}\label{poisson}
 \nabla^2\Phi=4\pi G\rho,
\end{equation}
 where $\rho$ is related to the mass of the simulation, $M$, and its
phase-space probability density, $f$, by
 \begin{equation}\label{getrho}
\rho(\b x)=M\int\d^3\b v\,f(\b x,\b v).
\end{equation}
 The integral in equation (\ref{getrho}) is evaluated by Monte-Carlo
sampling of velocity space. That is, one exploits the theorem that for a
wide range of functions $g$ we have
 \begin{equation}
\int\d z\,g(z)=\lim_{N\to\infty}{1\over N}\sum_{i=1}^Ng(z_i)/f_{\rm s}(z_i),
\end{equation}
 where the $z_i$ are $N$ points distributed through the domain of
integration with density $f_{\rm s}(z)$, the latter being normalized such
that $\int\d z\,f_{\rm s}=1$. We define a function $W_k(\b x)$ such that
outside the $k$th cell it vanishes, and its integral over the cell equals
unity.  Then we express the mean density in the $k$th cell as 
 \begin{eqnarray}\label{giverhok}
\rho_k&=&M\int\d^3\b x\d^3\b v\,W_k(\b x)f(\b x,\b v)\\
&=&\lim_{N\to\infty}{M\over N}\sum_{i=1}^NW_k(\b x_i)
{f(\b x_i,\b v_i)\over f_{\rm s}(\b x_i,\b v_i)}\nonumber
\end{eqnarray}
 In a conventional $N$-body simulation, the initial conditions of the
particles are chosen with probability density $f$, so $f_{\rm s}=f$
initially.  Since $f$ and $f_{\rm s}$ are constant along orbits, the two
functions remain equal, and the sum in equation (\ref{giverhok}) reduces to
the weighted number of particles in the $k$th cell:
 \begin{equation}\label{countit}
\rho_k=\lim_{N\to\infty}{M\over N}\sum_{i=1}^NW_k(\b x_i).
\end{equation}

\subsection{Cosmological $N$-body simulations}

There is usually a significant difference between a cosmological $N$-body
simulation and the conventional paradigm just presented in that in these
simulations the initial conditions do {\em not\/} randomly sample phase
space with probability density $f$. The standard procedure is to place the
particles at rest at the nodes of a regular lattice, and then to displace
them slightly in position and velocity according to the Zel'dovich
approximation (Efstathiou \ea 1985). In these circumstances, the
density is given by the Jacobian of the transformation from Lagrangian to
Eulerian coordinates:
 \begin{equation}\label{LEeq}
\rho(\b x)=\rho_0{\p(\b q)\over\p(\b x)},
\end{equation}
 where $\rho_0$ is the mean cosmic density and 
$\b q$ is the Lagrangian coordinate. Consequently, the particles are
at all times on a uniform lattice in $\b q$-space. If the density has the
band-limited form
 \begin{equation}
\rho=\sum_{|\b k|<K}\hat\rho_\b k\e^{\i\b k.\b x},
\end{equation}
 then it is straightforward to show that one can exactly recover $3N$
Fourier amplitudes from the coordinates of $N$ particles that are
distributed on a uniform lattice in $\b q$ and a slightly distorted lattice
in $\b x$ (Appendix A). By contrast, if we randomly sampled the density
field $\rho(\b x)$ with $N$ particles, and then tried to recover $\rho$ from
the particle coordinates, the Fourier coefficients of the recovered density
would be significantly in error for larger values of $|\b k|$.

Once particles have moved far from their initial positions $\b x=\b q$,
equation (\ref{LEeq}) ceases to be useful. We then argue that at very high
redshift, when the co-moving distribution function was $f(\b x,\b
v)=f_0\delta(\b v)$, with $f_0$ a constant, the particles uniformly sampled
$f$ in the sense that they lay at rest on a uniform grid in $\b x$. The
constancy of $f$ along orbits implies that the particles always uniformly
sample the part of phase space in which $f\ne0$, and we can estimate $\rho$
from equation (\ref{countit}) as in a conventional $N$-body simulation.

The fact that we have two fundamentally different ways of determining
density in a cosmological simulation is generally obscured because 
Poisson's equation is side-stepped in favour of
Poisson's integral for
the gravitational force,
 \begin{equation}\label{directf}
\b F(\b x)=-GM\int\d^3\b x'\d^3\b v'\,
f(\b x',\b v'){\b x-\b x'\over|\b x-\b x'|^{3/2}}.
\end{equation}
 It is now assumed without detailed enquiry, that the particles are
distributed with probability density $f$, so that the integral can be
approximated as
\begin{equation}
\b F(\b x)={GM\over N}\sum_{i=1}^NG(\b x-\b x_i),
\end{equation}
 where in a naive application of the theory of Monte-Carlo integration the
Green's function $G$ would be $G(\b x)=-\b x/|\b x|^{3/2}$. In practice a
more complex form of $G$ is used because the integrand is singular at $\b
x=\b x'$ and one wishes to avoid a large variance in the
estimates  of the integral yielded by different random
distributions of points. Dehnen (2001) discusses the merits of various
possible forms of $G$ that all satisfy the general requirement
 \begin{equation}
G(\b x)\to\cases{
-{\displaystyle{\b x\over|\b x|^{3/2}}}&for $|\b x|$ large,\cr
0&for $|\b x|\to0$.}
\end{equation}
 Let $\epsilon$ be the `softening' radius within which $G$ deviates
significantly from the inverse-square law. Cosmological simulators
generally consider that $\epsilon$ should be as small as it can be,
and in any case less than the inter-particle separation in the initial
state. To our knowledge the correctness of this proposition has not
been demonstrated in the literature. On the contrary, Knebe \ea (2000)
have shown that great care has to be taken when choosing the softening length
if unphysical two-particle scattering events are to be avoided. The
discussion above shows that there are really two questions, namely,
what value of $\epsilon$ yields the best approximation to the forces
(i) at early times, when equation (\ref{LEeq}) is valid, and (ii) in
the virialized regime when equation (\ref{countit}) applies?  We have
seen that in the first regime the density field can be determined
right down to the scale of the inter-particle separation. Hence, small
values of $\epsilon$ are appropriate in this regime. In the virialized
regime, the fractional uncertainty in the density on the scale of a
cell that contains $n$ particles is $\sim n^{-1/2}$. Hence, in this
regime $\epsilon$ should exceed the interparticle separation by some
factor. We determine appropriate values of $\epsilon$ below.

\section{MLAPM's Poisson solver}\label{multigrid}

 MLAPM does not use a Green's function to sum inter-particle forces, but
estimates the density on an adaptive grid and then employs a
finite-difference approximation to solve Poisson's equation subject to
periodic boundary conditions.
The entire computational domain is covered by a hierarchy of `domain grids'
that have $2^n$ cells on a side. The finest domain grid has at least as many cells as
there are particles in the simulation, and the coarsest grid has 2 cells on a
side. If the density in any cell is found to exceed a density threshold,
which corresponds to $\rho_{\rm ref}$ 1 to 8 particles per cell, the cell is
subdivided as described below.  Cells obtained on this subdivision can be
further subdivided, and so on indefinitely. This sub-division process, which
can generate grids of arbitrary geometry, is described in more detail in
Section \ref{sec:refine}.

To define and navigate such complex grids, several data structures are
required, which we now describe. The general scheme closely follows the
precepts of Brandt (1977). Functions are provided both for the creation and
destruction of these structures.

With each cell we associate a data structure called a  `node', which stores
the values for the centre of the cell of dynamically interesting quantities:

\begin{tabular}{lr}\hline
NODE & 
 \begin{tabular}{l}
  $\circ$ density \\
  $\circ$ potential \\
  $\circ$ forces \\
  $\circ$ pointer to first particle \\
 \end{tabular} \\ \hline
\end{tabular}
\\

\noindent
Since there will be more nodes than particles, they need to be defined
in a way that minimizes memory requirements. Moreover, so far as
possible, we arrange for nodes that are adjacent physically to occupy
adjacent locations in computer memory. This has the dual advantage of
minimizing cache misses and of enabling neighbours to be found by
incrementing or decrementing pointers.  Hence we do not follow
Kravtsov~\ea~(1997) in arranging nodes as fully-threaded oct-trees.
Instead we gather nodes into $x$QUADs.  An $x$QUAD is a line of nodes
that follow each other parallel to the $x$-axis. With it we associate
these numbers

\begin{tabular}{lr}\hline
$x$QUAD & 
 \begin{tabular}{l}
  $\circ$ pointer to first node \\
  $\circ$ $x$ coordinate of the first node \\
  $\circ$ number of nodes \\
  $\circ$ pointer to next $x$QUAD \\
 \end{tabular} \\ \hline
\end{tabular}
\\

\noindent
Since the memory for the
nodes described by this QUAD is allocated as one block, this information
is sufficient to access directly any node in the QUAD and to determine its
$x$ coordinate. The pointer to the next $x$QUAD 
similarly enables one to reach nodes further down the axis in a few steps.

   \begin{figure}
		\centerline{\psfig{file=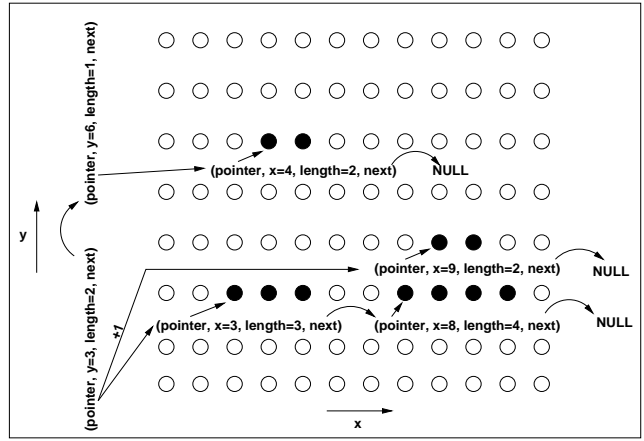,width=\hsize}}
      \caption{QUAD structured grid used within MLAPM sketched for 
               two dimensions. Circles mark  nodes, open ones being virtual.
			   QUADs are indicated by lists in brackets.}
      \label{QUAD}
    \end{figure}

Just as nodes are gathered into $x$QUADs, so $x$QUADs are gathered into
$y$QUADs.  Thus a $y$QUAD is a series of contiguous $x$QUADs and gives one
access to a plane\footnote{Brandt calls a $y$QUAD a CQUAD.} of nodes. With a
$y$QUAD we associate these numbers 

\begin{tabular}{lr}\hline
 $y$QUAD &  
 \begin{tabular}{l}
  $\circ$ pointer to first $x$QUAD\\
  $\circ$ $y$ coordinate of first $x$QUAD\\
  $\circ$ length of $y$QUAD\\
  $\circ$ pointer to next $y$QUAD\\
 \end{tabular} \\ \hline
\end{tabular}
\\

\noindent
A $z$QUAD is a similar linked list of $y$QUADs, so it contains these numbers

\begin{tabular}{lr}\hline
 $z$QUAD &  
 \begin{tabular}{l}
  $\circ$ pointer to first $y$QUAD\\
  $\circ$ $z$ coordinate of first $y$QUAD\\
  $\circ$ length of $z$QUAD\\
  $\circ$ pointer to next $z$QUAD\\
 \end{tabular} \\ \hline
\end{tabular}
\\

\noindent
Fig.~\ref{QUAD} indicates how a two-dimensional, adaptive grid is
organized using QUAD's.  All (virtual) nodes of a grid are shown, with
the nodes in use (refined region) represented by filled circles.
Memory is assigned only for these nodes (and the supporting QUAD
structures).  As soon as a node is encountered that does not need to
be refined, the $x$QUAD stops and its `next'-pointer is set to the
next $x$QUAD; if this is the last $x$QUAD, the pointer is set to
NULL. The same scheme applies to the relation between $x$QUAD's and
$y$QUAD's, and to the relation between $y$QUADs and $z$QUADs.  In
particular, when a series of $x$QUADs is contiguous in the sense that
there is at least one $x$QUAD for every value of $y$ in some range,
the storage for the $x$QUADs with the smallest $x$ coordinates at each
$y$ is allocated in a block. Similarly, storage for contiguous
$y$QUADs with the smallest $y$ coordinates at given $z$ is allocated
in a block.

Computation of the forces involves several sweeps through the nodes. In each
such sweep one loops through the linked list of all $z$QUADs to locate each
$y$QUAD, and within each $y$QUAD one runs through the list of $x$QUADs, and
within each $x$QUAD one runs through the list of nodes.  Consequently, when
referencing a node one always knows which $x$QUAD, $y$QUAD and $z$QUAD it
lies in. This information and the coherent storage of adjacent $x$ and
$y$QUADS allows one to find neighbours as follows. For example, suppose we
want to find the neighbour that has $y$ smaller by a grid spacing. Then we
decrement by one the current value of the pointer in the loop over $x$QUADs
to locate the $x$QUAD nearest the $y$-axis at the required value of $y$.
Then we loop over the list of $x$QUADs at whose head this QUAD stands, until
we find the $x$QUAD that contains the neighbour we are seeking.

The highest-level structure in MLAPM is a GRID. This gathers together a
variety of information about a particular level of refinement:\\

\begin{tabular}{lr}\hline
GRID &
 \begin{tabular}{l}
  $\circ$ pointer to first $z$QUAD \\
  $\circ$ number of nodes per dimension \\
  $\circ$ distance between adjacent nodes\\
  $\circ$ critical density \\
  $\circ$ mass to density conversion factor\\
  $\circ$ residuals \\
  $\circ$ cosmic expansion factor \\
  $\circ$ ...
 \end{tabular} \\ \hline
\end{tabular}\\

\noindent
The crucial entries in this structure are the pointer to the first
$z$QUAD and the number of  (virtual) nodes. However, additional useful book-keeping
data is stored here, such as the grid spacing, and the critical density
for refinement. The roles of several of these quantities will become clear
later. 

The data structure associated with a particle is this

\begin{tabular}{lr} \hline
PARTICLE &
 \begin{tabular}{l}
  $\circ$ position \\
  $\circ$ momentum \\
  $\circ$ pointer to next particle \\
 \end{tabular} \\ \hline
\end{tabular}
\\

\noindent Each particle is assigned to a node, usually the finest node that
contains it. The list of a nodes's particles is maintained as a standard
linked list.  These linked lists are sorted with respect to the
$x$-coordinate.

\subsection{Memory Requirements}\label{sec:memory}

Since cosmological simulations are often limited by available
memory rather than processor time, it is important to keep track of memory
requirements. Here we assume that each floating-point number requires 1 word
of storage (usually 4 bytes) and each pointer 2 words.

The storage requirement is dominated by particles, which require 8 words
each, and nodes, which require 7 words each. If the finest domain grid has
$2^L$ nodes on a side, between them the domain grids contain
$2^{3L}+\cdots+2^6=\fracj87(2^{3L}-8)$ nodes and thus require almost exactly
the same number of words ($ 2^{3(L+1)}$) as do $2^{3L}$ particles.

Each QUAD requires just 6 words of storage and there are very many fewer
quads than nodes, so their storage requirement is unimportant.

\subsection{Refinements}\label{sec:refine}

\begin{figure}
      \centerline{\psfig{file=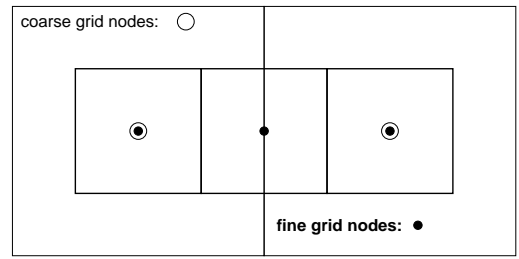,width=.8\hsize}}
      \caption{Fine-grid cells often overlap more than one coarse-grid cell.
	  Consequently, the fine-grid node at the centre may owe its existence to
	  either of the two coarse-grid nodes exceeding the density threshold.}
      \label{cells}
    \end{figure}

A node is refined if its density exceeds a predetermined threshold
that varies from grid to grid, and de-refined whenever it falls below
that value. However, around each high-density region some
additional nodes are refined, to provide a `buffer zone'.
These buffer zones ensure that the resolution of the grid changes only
gradually even if the density is discontinuous.
In detail, a node is refined if either its density, or the density of any of
the 26 surrounding nodes exceeds the density threshold. Consequently, as
MLAPM marches through the grid deciding whether to refine nodes, it is
continually testing the density of nodes that lie ahead of its current
position, since the current node must be refined if any of them lie above
the density threshold.  Careful programming is required to avoid wasting
time by testing nodes twice. Notice that a refined node such as that shown
in the centre of Fig.~\ref{cells} can be called into existence by virtue of
the coarse node to its right or to its left exceeding the density threshold,
so we do not speak of `parent' and `child' nodes.

An important difference between our refinement scheme and that of the ART
code, is that some of our refined nodes are cospatial with coarse nodes (see
Fig.~\ref{transfig} below), whereas in the ART code all refined nodes are
symmetrically distributed within the parent coarse node. Our refinement
scheme is the natural one to adopt if one is simply solving partial
differential equations. When particles are involved, it does lead to
additional complexity, however, because with our scheme refined nodes that
are not cospatial with coarse nodes have cells that overlap the cells of
more than one coarse node -- see Fig.~\ref{cells}.

The edges of refinements always include cospatial nodes of the parent grid
(e.g., Fig.~\ref{transfig} below).
Nodes that lie on the boundary of a refinement have a different role from
ones in the interior. First they carry the boundary conditions subject to
which Poisson's equation is solved in the interior of the refined grid. That
is, the potential on a refinement's boundary nodes is obtained by
interpolation from the embedding coarse grid and held constant as the
potential at interior points is adjusted towards a solution of Poisson's
equation as described in Section \ref{sec:relax}.  The second role of
boundary nodes is to carry values used in the determination of the forces on
particles in the refinement -- the determination of these forces involves
both numerical differentiation and interpolation.

\subsection{Particle assignment}\label{sec:partassign}

Generally, each particle is placed in the linked list of the finest node
within whose cell it lies. Exceptions to this rule occur when a particle
enters a refinement during a call to STEP (see Section \ref{SECstepping}) and on
the boundaries of refinements, where refined nodes exist only
to provide values of the potential and forces. These nodes do not acquire
particles.

After testing the nearest-grid-point (NGP), cloud-in-cell (CIC) and
triangular-shaped-cloud (TSC) mass-assignment schemes (Hockney \&
Eastwood 1988) we adopted the TSC mass-assignment scheme. In both the
CIC and TSC schemes a particle contributes to the density in more than
one node. Particular care has to be exercised at the edges of
refinements if the integral of the density is to equal the total mass
of the particles.

A particle in the interior of a refinement only contributes to the density
at refined nodes. When the density at cospatial coarse nodes is required, it
is set equal to a weighted mean of the densities on a number of nearby fine
nodes. Brandt (1977) calls this operation of taking a weighted mean
`restriction'. The operator that accomplishes it has to be matched to the
mass-assignment scheme, so that one obtains the same coarse-grid densities
by restriction from a fine grid as one would have obtained if there had
been no refinement and particles had been assigned to the coarse grid.

The restriction operator is also matched to an interpolation operator that
is used to estimate quantities on a fine grid from their values on the
embedding coarse grid. Brandt calls this the `prolongation' operator. The
matching is such that if values are prolonged from coarse to fine and then
restricted back to the coarse grid, they do not change.

Intricate book-keeping is required when particles are transferred
between grids on the creation of a refinement -- some details are given in
Appendix B.

\subsection{Relaxation Procedure}\label{sec:relax}

Poisson's equation is solved using a variant of the multigrid technique
(Brandt 1977; Press et al. 1992). In essence one relaxes a trial
potential to an approximate solution of Poisson's equation 
by repeatedly updating the potential according to
\begin{equation} \label{sweep}
 \begin{array}{lcl}
  \Phi_{i,j,k} & = &\fracj16 
                    (\Phi_{i+1,j,k}+\Phi_{i-1,j,k}+\Phi_{i,j+1,k}+ \\
               &   & \qquad \Phi_{i,j-1,k}+\Phi_{i,j,k+1}+\Phi_{i,j,k-1}-
               \rho_{i,j,k}\Delta^2) \ ,
 \end{array}
\end{equation}
 where $\Delta$ is the grid spacing. There are several possible orderings of
the points $(i,j,k)$ at which these updates are made. We
use `red-black' ordering, so called because it involves first updating
$\Phi$ on every other node on the grid, as on the red squares of a chess
board, and then updating the other half of the nodes, equivalent to the
black squares on a chess board.

This algorithm rapidly eliminates errors in the trial potential that
fluctuate on the scale of the grid, but eliminates errors with longer-range
fluctuations much more slowly. The multigrid technique involves using  a
coarser grid to seek a correction in the event that convergence is slow. 

We start the iteration process on the finest domain grid, usually with the
potential from the last time step. This is iterated to convergence, if
necessary with use of the coarser grids. (On the coarsest, $2^3$, grid the
difference equations are solved analytically.) Once we have a solution on
the domain grid, we prolong it to any refinements and iterate on the
refinements. Each refinement poses an independent boundary-value problem.
In general these problems cannot be posed on a coarser grid because the
boundary includes nodes not present on the coarser grid. Hence we are
obliged to iterate to convergence on the refinements alone. Fortunately, the
trial potential only deviates from the true one on the finest scales because
it is obtained by prolongation of a coarse-grid solution of the same
problem. So convergence is in practice rapid. Any further refinements are
handled in the same way.

The potential on any grid is deemed to have converged when the residual
 \begin{equation}\label{residual}
 e = \nabla^2 \Phi - \rho
\end{equation}
is smaller than a fraction, $\sim0.1$, of the
estimated truncation error $\tau$. We estimate the latter as
 \begin{equation}\label{truncation}
 \tau = \wp\left[\nabla^2 (\Re \Phi)\right] - (\nabla^2 \Phi),
\end{equation}
 where $\wp$ and $\Re$ are the prolongation and restriction operators,
respectively. Thus, $\tau$ is essentially the difference between evaluating
the Laplacian operator on the next coarser grid and on the current grid.

Forces at each node are evaluated from centred differences of the potential
and propagated to the locations of particles by the TSC scheme to ensure
exact momentum conservation within any given refinement (Hockney~\&~Eastwood
1988).  As in any code with adaptive softening, momentum is not precisely
conserved when refinements are used. In Section 5.1.3 below we quantify this
problem in two specimen configurations.

\begin{figure}
\psfig{file=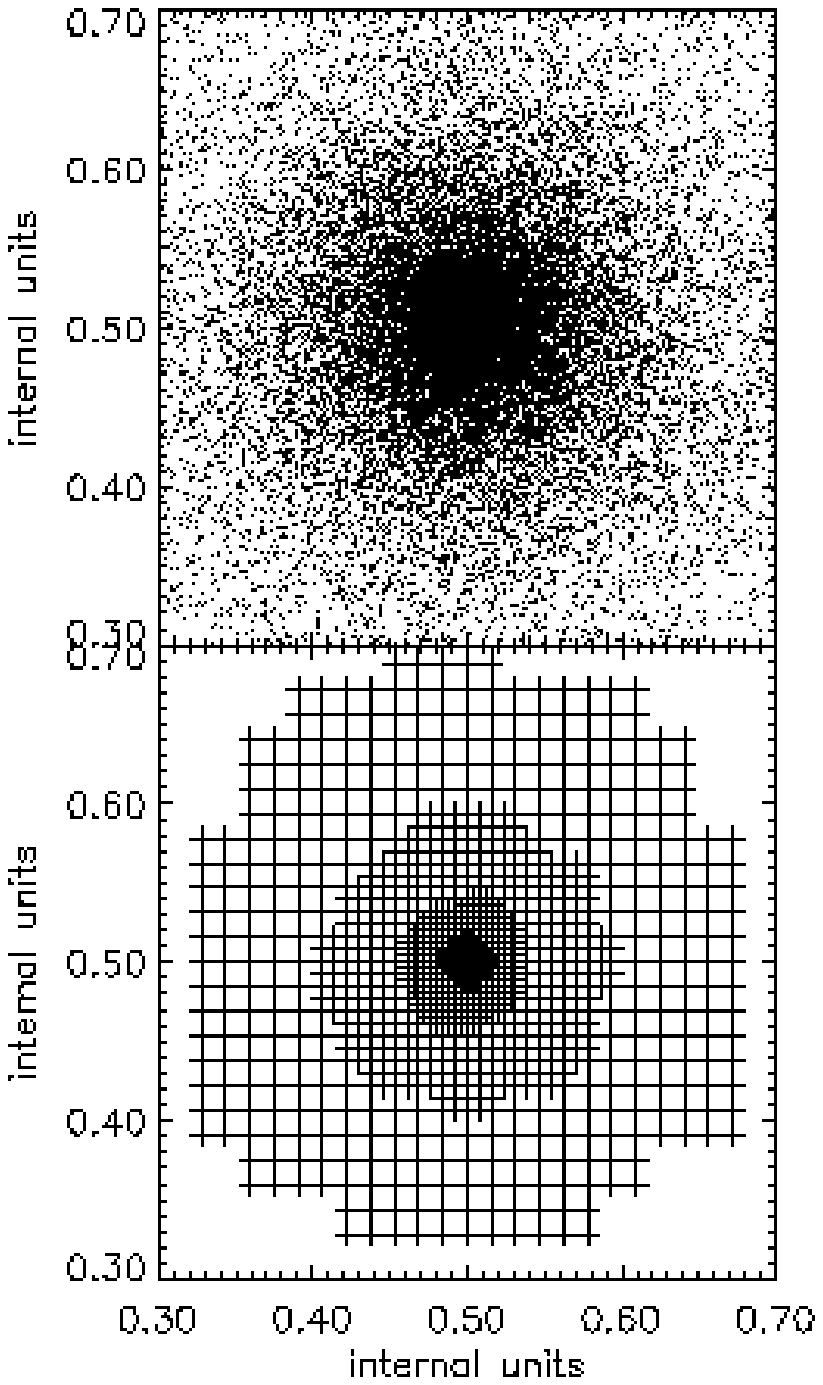,width=\hsize}
\caption{Refinement structure for Hernquist model sampled by $32^3$ particles
         and using $\rho_{\rm ref} = 8$ particles per node.\label{mlapm_ref}}
\end{figure}

 \section{Performance of the Poisson solver}
 \label{SECperformance}
The writers of $N$-body codes traditionally check the accuracy of
their Poisson solver by using it to calculate the force between two
point masses at various separations. In our view this test is
misguided because a Poisson solver that is adapted to the solution of
the collisionless Boltzmann equation should not return the force
between point particles. At some level this fact is widely recognized
in that a `softened' interparticle force is aimed at, but isotropy of
the interparticle force is still considered desirable.  A Poisson
solver for collisionless dynamics is concerned with finding the forces
generated by {\em smooth\/} mass distributions. A single particle
corresponds to a mass distribution that is unresolved on any smoothing
scale, and thus one that falls outside its remit.  Presented with this
ill-posed problem, the best it can do is to assume that the density is
non-zero in the cells around the particle, and zero
elsewhere. Inevitably, this mass distribution reflects the geometry of
the code's cells, and it will not generate an isotropic gravitational
field.

When testing a Poisson solver we should check its ability to recover
the potential of density distributions of the type that it will
encounter in the field.  We have tested our code by comparing with
analytic results the forces it generates for (a) a Hernquist model,
and (b) a plane wave.

\begin{figure*}
\centerline{\psfig{file=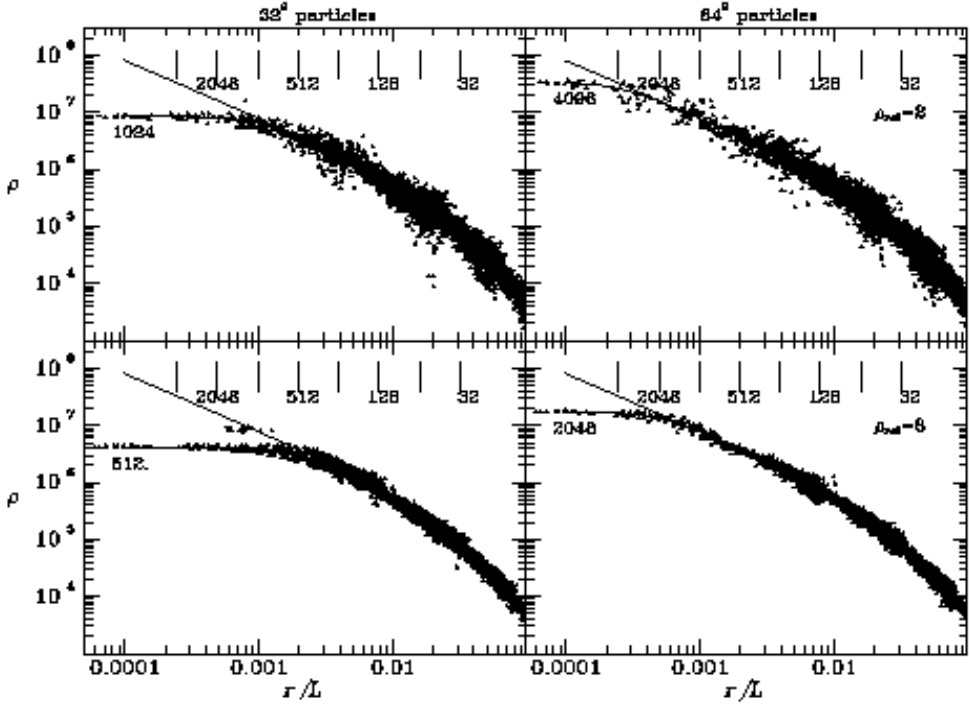,width=.8\hsize}}
\caption{Recovery of the density profile of a Hernquist model from
particle positions.  For the left panels $32^3$ particles sampled a
Hernquist profile, shown as the upper curve, with scale radius
$\fracj1{16}$ of the box size and outer cutoff half the box size,
$L$. For the right panels $64^3$ particles sampled the same
profile. For all panels the domain grid had 32 nodes on a side. For
the upper panels a node was refined if its density exceeded two
particles per node, while for the lower panels the refinement
threshold was eight particles per node. The tick marks along the top
show the sizes of cells of grids with 4098, 2048, \dots, nodes on a
side. The lower curves show the effect of doubly convolving the
Hernquist profile with the TSC mass-assignment
kernel. In the lower left panel, a small number of particles
lie above the main mass near $r/L=0.001$. 
This phenomenon reflects the creation of a small refinement
centred on the region of maximum density, which  Poisson noise has displaced
slightly from
the centre of the probability distribution ($r=0$). In most
realizations this feature is absent.\label{denstyfig} }
\end{figure*}

\subsection{Hernquist model}

We check the reliability of our refinement procedure
and investigate the origin of errors in the force, by
sampling a Hernquist model, in which the density varies with radius as
 \begin{equation}\label{HQST}
 \rho(r) =  {M r_0\over 2 \pi} {1\over r (r_0 + r)^3}.
\end{equation}
 The scale length $r_0$ of the Hernquist model was set equal to
${1\over16}$ of the box size, and we calculated the potential with a
domain grid $32$ nodes on a side. The model was truncated at a radius
of $16$ grid nodes and a uniform background density was added to make
the mean density within the box equal to a predetermined cosmic value;
in practice about 3/7 of all particles were associated with the
background. Our analytic calculations of the force do not include
contributions coming from outside the box, where the periodic boundary
conditions ensure that there are infinitely many other Hernquist
models.

\subsubsection{Refinement Hierarchy}

Fig.~\ref{mlapm_ref} gives a visual impression of our refinement
hierarchy at work by showing the distribution of particles according
to a Hernquist model sampled with $32^3$ particles, and the threshold
for refining nodes was set to $\rho_{\rm ref}=8$ particles per node. Refinements are
shown down to the level of $512^3$ (virtual) nodes. One can clearly
see how the grid structure adapts to the actual particle/density
distribution. Successively more accurate solutions to Poisson's
equation are achieved within regions of higher density, where better
force resolution is required to follow properly the particle dynamics.

\begin{figure}
\psfig{file=eps/ngrid.ps,width=\hsize}
\caption{The distribution of particles and nodes over grids in the realizations of a
Hernquist sphere shown in Fig.~\ref{denstyfig}. Full histograms: the numbers
of particles in each grid. Hatched histograms: the numbers of nodes in each
grid. The normalization $N_{\rm part}$ equals $32^3$ or $64^3$ rather than the actual
number of particles in the simulation, which is slightly larger, as
explained in the text.\label{ngridfig}}
\end{figure}

\begin{figure*}
\psfig{file=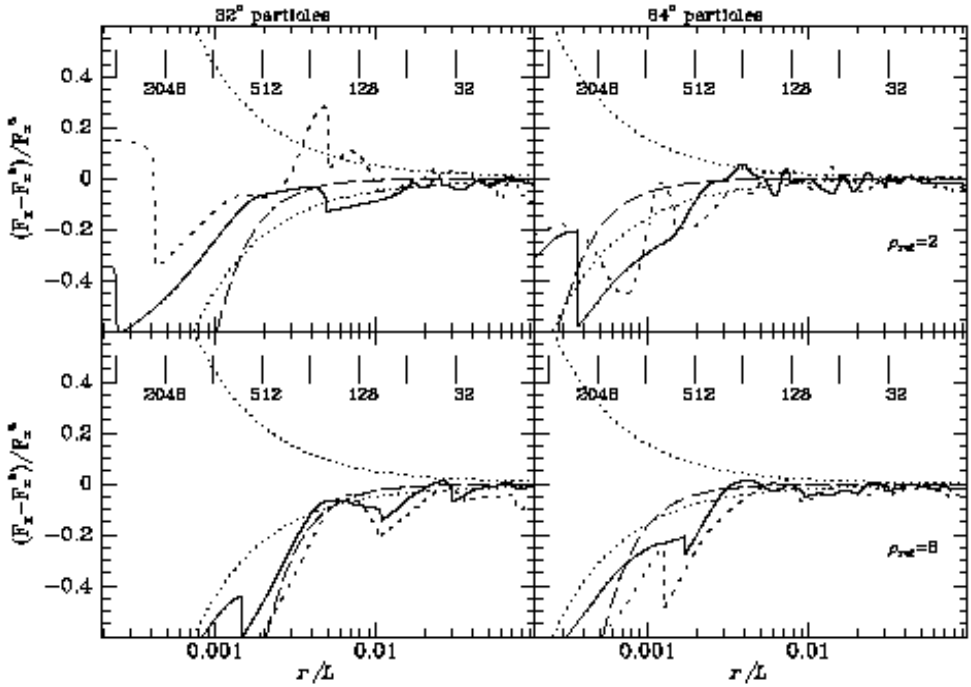,width=.8\hsize}
 \caption{For the Hernquist model described in the text, the fractional
difference between the values returned by MLAPM and obtained analytically
for the $x$ component of the force. Distance from the centre of the
sphere is plotted horizontally. The full curve is for values along the
$x$-axis, while the short dashed curve is for values along the 111 direction
relative to the axes. The upper panels are obtained when the critical
density for refinement, $\rho_{\rm ref}$, corresponds to 2 particles per
node, while the lower panels correspond to 8 particles per node.
\label{Hernfig}}
\end{figure*}

\subsubsection{Density estimates}

It is important to know how accurately one can recover the density
within a structure from the positions of particles that randomly
sample it. The standard theorem of Monte-Carlo integration states that
\begin{equation} \overline{\rho}(\b r)\equiv\int\d^3\b r'\,W(\b r-\b
r')\rho(\b r') =\lim_{N\to\infty}{M\over N}\sum_{\alpha=1}^N W(\b r-\b
r_\alpha),
\end{equation}
 where the $\b r_\alpha$ are positions distributed with probability density
proportional to $\rho(\b r)$. Applying this result to the case when $W(\b
r-\b r_\alpha)$ is the fraction of the mass of a particle at $\b r_\alpha$
that is assigned to a node at $\b r$, we see that in the limit of infinitely
many particles, the values of the density on the grid are not those of the
input density $\rho$ but its convolution $\overline{\rho}$ with the
mass-assignment kernel $W$. Moreover, if we use the same mass-assignment
scheme to interpolate these values back to positions that are not on the
grid, we recover
 \begin{equation}
\overline{\overline{\rho}}(\b r)\equiv\sum_{{\rm nodes}\ i}
W(\b r-\b r_i)\overline{\rho}(\b r_i),
\end{equation}
 which is a discrete approximation to the convolution of $\overline{\rho}$
with the mass-assignment kernel. Hence, we expect density values recovered
from the code to reflect not the input density but its double convolution
with the mass-assignment kernel. 

Fig.~\ref{denstyfig} shows that this expectation is borne out by showing
four attempts to recover the density of the Hernquist sphere from the
positions of either $32^3$ particles (left-hand panels) or $64^3$ particles
(right-hand panels). In each case the recovered densities scatter around the
result of doubly convolving the input density distribution with the TSC
kernel for a grid with from 512 to 4096 nodes on a side. Increasing the
particle number by a factor 8 causes finer grids to be generated, and thus
enables the model's $r^{-1}$ core to be traced further in. On the other
hand, the variance in the estimated densities is not decreased by an
increase in particle number.  The upper panels show the result of refining
nodes at a lower density threshold ($\rho_{\rm ref}=2$ particles per node)
than the lower ones ($\rho_{\rm ref}=8$ particles per node). The reduction
in variance and loss of resolution caused by an increase in the density
threshold are evident. Also evident in the lower right panel is the increase
in the variance as the edge of each grid is approached; at the outside of a
grid the number of particles per node is smallest, and the variance
correspondingly high.

Fig.~\ref{ngridfig} shows that lowering the critical density for refinement
from eight to two particles per node does increase the maximum spatial
resolution, but at considerable computational cost. Whereas with $\rho_{\rm
ref}=8$ the ratio $n_{\rm node}/n_{\rm part}=0.75$ (for $64^3$ particles),
this ratio rises to 3 when $\rho_{\rm ref}=2$.  A node has a greater
computational cost than a particle and it is less useful scientifically.
Resources spent on lowering $\rho_{\rm ref}$ would be better spent
increasing the number of particles. 

In all four realizations the vast majority of nodes belong to the grids with
less than 256 nodes on a side. Figs.~\ref{denstyfig} and \ref{Hernfig} below
show that on these scales little is gained by using a low value of
$\rho_{\rm ref}$ -- the gains from lowering $\rho_{\rm ref}$ are
concentrated at small radii and derive from grids that contain small numbers
of nodes and particles. In fact the numbers of nodes in the 4096$^3$ grid in
the top-right panels of Figs~\ref{denstyfig} and \ref{Hernfig} are so small
that they cannot be seen in Fig.~\ref{ngridfig}. These findings suggest that
significant gains in efficiency could be obtained by basing the refinement
criterion on the truncation error in the forces rather than on the density.
However, implementing this proposal is a job for the future.

\subsubsection{Force estimates} \label{SECforceshape}

Fig.~\ref{Hernfig} is similar to the Fig.~\ref{denstyfig} but for the
estimated gravitational field $\b F$ of the Hernquist model.  Again left
panels show results obtained with $32^3$ particles and right panels results
for $64^3$ particles, and the upper and lower panels are for $\rho_{\rm
crit}=2$ and 8 particles per node, respectively. In each panel the dotted
curves show the loci $y(r)=\pm[\overline{N}(r)]^{-1/2}$, where
$\overline{N}(r)$ is the expected number of particles interior to $r$. These
curves show the minimum variance from Poisson noise: the true variance will
be larger because density fluctuations are not constrained to be spherically
symmetric.  The full curve shows the difference between the analytic and
numerical values of $F_x$ as a function of radius along the $x$ axis, while
the short dashed curve shows the same quantity along the line $(1,1,1)$.
These two curves agree with one another to within the anticipated Poisson
errors, which shows that grid-generated anisotropy is not a problem.  The
long-dashed curves show the error expected because even in the limit of
infinitely many particles the mass-assignment scheme recovers not the true
density but its double convolution with the mass-assignment kernel. It is
evident that the variance and the bias in the forces are fully accounted for
by Poisson noise and smoothing by the mass-assignment kernel.

This conclusion is confirmed by a test in which the analytic value of the
density was placed on every node before solving for the forces: on grid
nodes the resulting forces agreed with the analytic ones to better than
$0.2$ percent at all points, and to better than $0.05$ percent further from
the centre than $2\Delta$ for the finest grid.

\begin{figure}
\psfig{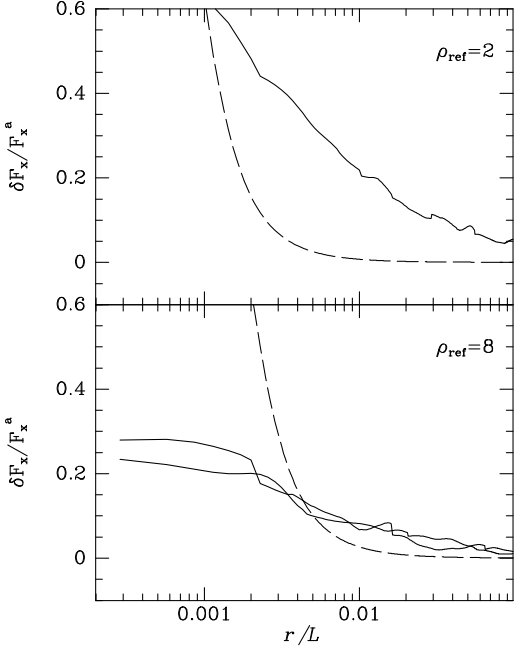}
\caption{Bias and variance  with two values of
$\rho_{\rm ref}$. The dashed curves shows the relative error in the force
of the Hernquist sphere that arises because the density is twice convolved
with the mass-assignment kernel. The full curves show the rms variation
in the force between different realizations of the system. All curves are
for the case of $32^3$ particles except the second full curve in the lower
panel, which is for $64^3$ particles. (It is on top at $r<0.001L$.)\label{varifig}}
\end{figure}

The discussion above can be summarized by saying that the errors in the
forces are dominated by uncertainty in the density. The latter is made up of
a systematic bias due to any unresolved core, and variance due to Poisson
noise.  Increasing the particle number at fixed $\rho_{\rm ref}$ decreases
the bias while holding the variance constant.  Increasing the threshold
density $\rho_{\rm ref}$ diminishes the variance and increases the bias.
Fig.~\ref{varifig} quantifies this last statement by plotting the bias in
the force from the left two panels of Fig.~\ref{Hernfig} as dashed curves,
and the rms variation in the force between different realizations of the
models as full curves. The latter decline outwards as the potential
fluctuations caused by local density fluctuations, which are always of order
$(\rho_{\rm ref})^{-1/2}$ times the local density, are increasingly swamped
by the barely changing mean inward pull of the model. This dilution of the
effects of density fluctuations is more marked in more massive systems, and
less marked in less massive ones. Since all halos start out as small
systems, we cannot rely on dilution of fluctuations to make our simulations
credible. We have to recognize from Fig.~\ref{varifig} that the uncertainty
in the forces can be reduced only by increasing $\rho_{\rm ref}$, and thus
reducing the simulation's spatial resolution. In particular, the
introduction of a Particle--Particle step to harden the inter-particle
forces at small separations would be analogous to lowering $\rho_{\rm ref}$
and therefore increasing the Poisson noise. We shall see in Section 5.2
below that our inter-particle force has a softening length of order
$2\Delta$, or about four times the inter-particle separation when $\rho_{\rm
ref}=8$.  Simulations with both \pcm\ and tree codes typically employ
softening lengths that are substantially smaller than the initial
inter-particle separation. Such small softening lengths are used because in
these codes the softening length is fixed in either physical or comoving
coordinates, so a small, and initially inappropriate value is required if
high-density regions are to be adequately resolved once they have collapsed
and virialized.  With our method the softening length automatically adapts
to some multiple of the local interparticle separation.

Fig.~\ref{varifig} suggests that the smallest permissible value of
$\rho_{\rm ref}$ is 8 particles per node, which restricts fluctuations in
forces near the centres of structure to the 20--30 percent level. The range
of radii over which we have a reasonable representation of the underlying
model, runs outwards roughly from the radius at which the bias falls below
the variance. From Fig.~\ref{varifig} we see that with $32^3$ particles the
range is $r>0.005L$, and in this range the forces are accurate to better
than $10$ percent. With more particles the range would have a smaller lower
limit, but the maximum uncertainty in the  forces would increase towards
$\sim25$ percent in the limit of very large $N$.

As explained at the end of  Section \ref{sec:relax}, the sum of all the
forces on the particles cannot be expected to vanish since our softening is
adaptive. Quantitatively,  for $64^3$ particles and $\rho_{\rm ref}=2$
particles per node, we find
\begin{equation}
\left|\sum_i\b F_i\right|=1.4\times10^{-4}\sum_i|\b F_i|.
\end{equation}
 When the same number of particles are distributed in a complex clustering
pattern that evolved from realistic cosmological simulation, the
coefficient in this equation was $4.7\times10^{-4}$. In the absence of
refinements, the coefficient was $2.7\times10^{-7}$, and thus zero to the
precision of a floating-point variable.

\subsection{Zel'dovich waves}

We now turn from virialized structures, to explore the performance 
of MLAPM before such structures form. Specifically, we compare the
forces it generates with analytic results for plane waves.

Let $\b r$ be Eulerian coordinates and $\b q$ Lagrangian coordinates for an
ensemble of particles that are uniformly distributed in $\b q$-space. Then
for $a(t)$ a suitable function of time that increases from zero to unity,
the mapping
 \begin{equation}\label{rtoq}
\b r=\b q+{a\b k\over k^2}\cos(\b k.\b q)
\end{equation}
 with $\b k$ a constant vector, generates a density field in $\b
r$-space that provides an exact solution for the development of a
plane-wave cosmological perturbation in a flat universe (Zel'dovich
1970).  The corresponding forces are readily obtained by
differentiating $\b r$ twice with respect to time.

\begin{figure*}
\centerline{\psfig{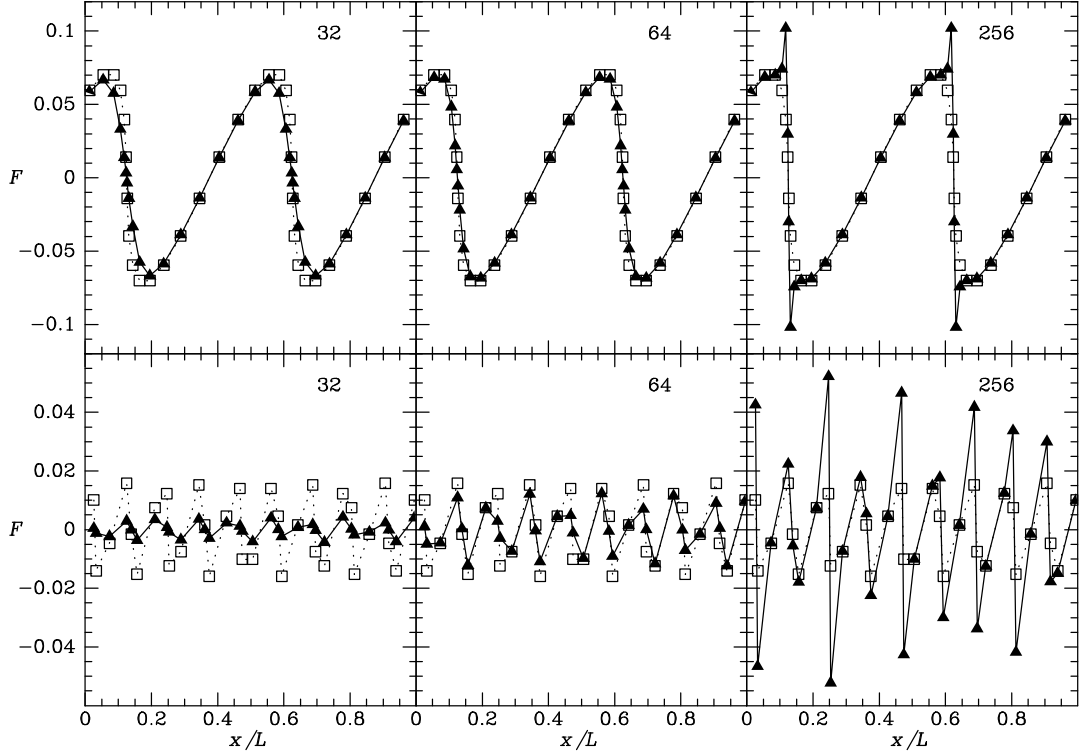}}
\caption{Forces from a pure PM code of Zel'dovich waves described by equation
(\ref{rtoq}) with $a=0.9$. In the top row $\b k=(4\pi/L,0,0)$
and in the bottom row $\b k=(18\pi/L,0,0)$. In every panel the wave is
sampled with $32^3$ particles, and the  grid has 32, 64 and 256 nodes on a
side as one runs from left to right. Analytic forces are marked by squares
and numerical ones by triangles.\label{PMgrid}}
\end{figure*}

\begin{figure*}
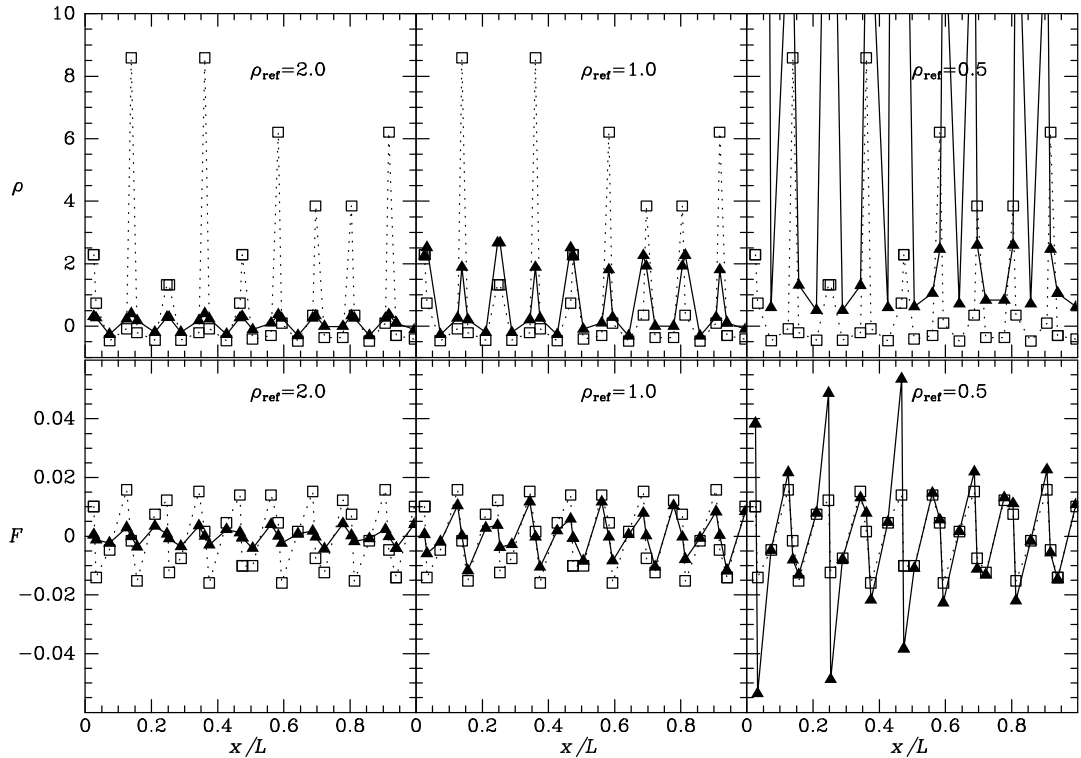

\centerline{\psfig{file=eps/DWdensity.ps,width=.8\hsize}}
\centerline{\psfig{file=eps/mlapm2.ps,width=.8\hsize}}
\caption{The density (top) and  forces (bottom) of a  
Zel'dovich wave with $\b k=(18\pi/L,0,0)$ recovered by MLAPM from the
positions of $32^3$ particles. The domain grid has 32 nodes on a
side. As in Fig.~\ref{PMgrid}, analytic values are marked by squares
and numerical ones by triangles.\label{MLAPMfig}}
\end{figure*}

Fig.~\ref{PMgrid} explores the ability of a simple PM code to recover the
forces generated by Zel'dovich waves with two values of $k$ and $a=0.9$.  In
every panel, the forces are recovered from the positions of $32^3$
particles. As one passes from left to right the number of nodes on a side of
the grid rises from 32 to 256. With as many nodes as particles the forces
are slightly in error for the longer wave and seriously in error for the
shorter one. When there are eight times as many nodes as particles, the
forces are reasonably accurate for both waves. With yet larger numbers of
nodes unphysical force spikes either side of the plane on which the wave
will break. One may readily demonstrate that these spikes arise because
particles approach each other very closely as the wave breaks, and with a
hard particle-particle interaction the overall force on a particle can be
dominated by the contribution from a single neighbour. In the case of the
shorter wave, the unphysical spikes make nonsense of the returned potential.
This experiment nicely demonstrates the importance of tuning the softening
of the potential to the resolution limit that is inherent in the number of
particles. 

\begin{figure*}
\centerline{\psfig{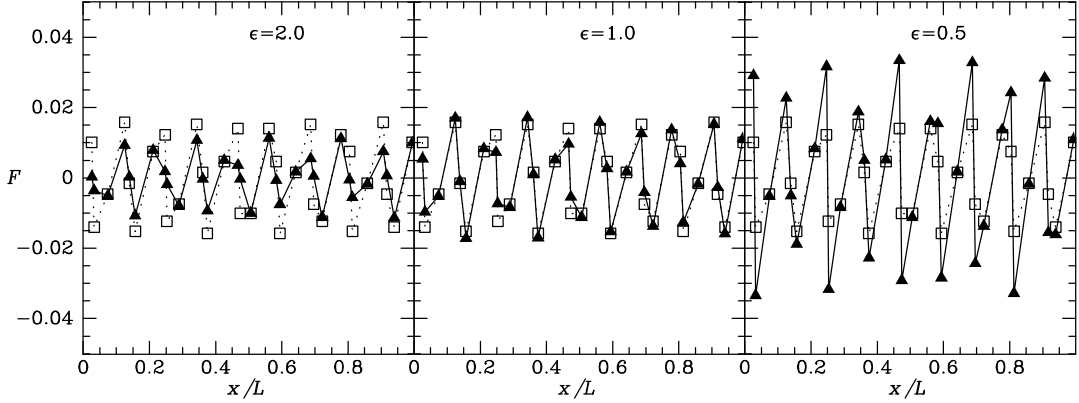}}
\caption{As Fig.~\ref{MLAPMfig} but showing forces recovered by Couchman's A\pcm\
code using a grid with as many ($32^3$) nodes as particles for three values of
the force softening.\label{AP3Mfig} }
\end{figure*}

Fig.~\ref{MLAPMfig} shows the density (top) and forces (bottom) that MLAPM
generates with $32^3$ particles, a domain grid that has 32 nodes on a side
and three values of the threshold density $\rho_{\rm ref}$. The most
accurate forces are obtained with $\rho_{\rm ref}=1$ particle per node.
Lower values again generate unphysical force spikes. Fig.~\ref{AP3Mfig}
shows that an A\pcm\ code also generates force spikes if the softening
parameter is smaller than the inter-particle separation.

To produce the results shown in Fig.~\ref{MLAPMfig} for $\rho_{\rm ref}=1$
particle per node, MLAPM refines most nodes of the domain grid once and none
twice. Consequently, there are $7.4$ nodes per particle, only slightly less
than if we had started with a domain grid eight times larger and $\rho_{\rm
ref}$ chosen to avoid refinement. We therefore have two strategies for
obtaining adequate resolution in unvirialized regions. In one strategy, the
domain grid has as many nodes as there are particles but we set $\rho_{\rm
ref}$ on the domain grid to a small enough value that it is essentially all
refined. Strictly we should ensure that the domain grid remains refined even
in voids until the density has fallen to $\sim\overline{\rho}/8$, and this
requires $\rho_{\rm ref}\simeq0.25$ particles per node. In practice such
small values of $\rho_{\rm ref}$ will be useful only at very late stages of
a simulation, because the second strategy is more economical so long as the
value of $\rho_{\rm ref}$ chosen under the first causes the whole domain
grid to be refined. In the second strategy, one starts with a domain grid
that has eight times as many nodes as particles, and sets $\rho_{\rm ref}=8$
on it because it provides adequate resolution until virialized structures
form. With many PM and \pcm\ codes, including the ART code, it is standard
practice (but not mandatory) to use such a large domain grid. In certain
circumstances this second strategy may be impossible on a given machine for
a given number of particles. Then the first strategy can be adopted with
$\rho_{\rm ref}$ set to the lowest value that is compatible with the
available hardware. So long as $\rho_{\rm ref}$ is comparable to or smaller
than unity, our experiments suggest that the correlation function and mass
functions obtained differ insignificantly from those obtained with the
second strategy (see Figs.~15 to 17 below).

Why is the optimal value of $\rho_{\rm ref}$ for Zel'dovich waves so much
lower than that appropriate for virialized structures? Why are Zel'dovich
waves best represented when 7/8 of the nodes are empty, and the remainder
contain only one particle?  There are two points to consider. (i) The TSC
mass-assignment algorithm distributes the mass of a particle over 27 nodes,
so a node may be empty and yet have non-zero density. (ii) A distribution of
particles placed on a Zel'dovich distorted grid differs markedly from the
particle distribution of a virialized body in that its underlying density
field {\em is\/} uniquely defined by the particles (Appendix A).
Hence, at early times the density field
in a cosmological simulation is defined up to the scale of the
inter-particle separation.  Since the matter distribution is represented by
particles, there is a great deal of artificial power on smaller scales, but
this power is rather cleanly separated from the lower-frequency power that
represents real cosmic fluctuations. As density gradients steepen
gravitationally, this separation becomes less clean, and it breaks down
completely with the formation of caustics and virialized structures.
Consequently, in virialized regions the density field is dominated by
Poisson noise at the scale of the inter-particle separation.

\section{Integrating the equations of motion} \label{SECstepping}

We now turn from the Poisson solver to consideration of how particles are
moved. 

\subsection{Time-stepping}

The Lagrangian for motion in comoving coordinates is
\begin{equation}
 {\mathcal L} = \fracj12 a^2 \dot{x}^2 - { \Phi\over a},
\end{equation}
 so the canonical momentum is
\begin{equation}
p = a^2 \dot{x},
\end{equation}
 and the Hamiltonian is
 \begin{equation}\label{spHamil}
 {\mathcal H} = {p^2 \over2 a^2} + {\Phi\over a} \ .
\end{equation}
 Hamilton's equations are therefore
 \begin{equation}\label{He}
 \begin{array}{lcc}
  \displaystyle {\d x\over\d t} & = &\displaystyle  {p\over a^2} \\ 
\\
  \displaystyle {\d p\over\d t} & = &\displaystyle  -{\nabla \Phi\over a} \ .
 \end{array}
\end{equation}
 We integrate these equations with a minor variant of the usual symplectic scheme of
second-order accuracy
\begin{eqnarray}\label{stepping}
  x_{n+1/2} & = & x_n + p_n
                 \int_t^{t+\Delta t/2}{\d t\over a^2} \nonumber\\
  p_{n+1}  & = & p_n - \nabla\Phi(x_{n+1/2})\int_t^{t+\Delta t}
                 {\d t\over a}\\
  x_{n+1}  & = & x_{n+1/2} + p_{n+1}
              \int_{t+\Delta t/2}^{t+\Delta t}{\d t \over a^2},  \nonumber
\end{eqnarray}
 where the integrals can be evaluated analytically because they depend only
on the cosmology.
The implementation of multiple timesteps described below requires that
positions and momenta be synchronized at the start and end of each timestep,
so we do not form the standard leapfrog scheme by combining the drift steps
that start and finish the above sequence of updates.

\begin{figure}
\centerline{\psfig{file=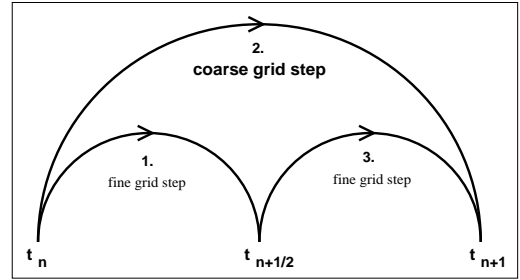,width=.8\hsize}}
      \caption{The principle of the recursive time-stepping scheme in the
	  case that part of the grid being operated on requires refinement. The
	  fine-grid steps 1 and 3 are accomplished by calling the full stepping
	  routine again and will typically involve further grid refinements. The
	  coarse-grid step 2 involves updating particles not previously moved
	  with the forces calculated with all particles advanced to the
	  half-time point.
      }
      \label{multistep}
    \end{figure}

On finer grids forces tend to be larger, and the time it takes a
particle to cross a cell is shorter. Hence shorter timesteps are
appropriate for finer grids.  Our time-stepping routine, STEP, is
called recursively. It takes as arguments a grid, $G$, and a time
interval, $\Delta$. STEP starts by asking whether any part of the grid
$G$ should be refined. If the answer is `no' it uses equations
(\ref{stepping}) to advance the particles on $G$ by $\Delta$ and then
returns. If refinement is in order, a refined grid $G'$ is created and
STEP($\Delta/2,G'$) called. That is, the particles on $G'$ are
advanced by $\Delta/2$ with the particles on $G$ still at the initial
time. Once $G'$ has been advanced in this way, STEP uses equations
(\ref{stepping}) to advance the particles still on $G$ by $\Delta$,
and then calls STEP($\Delta/2,G'$). STEP then erases $G'$ and
returns. This scheme is sketched by the following pseudo
$C$-code:

\begin{verbatim}
Step(dt, CurrentGrid){
  NewGrid = Refine(CurrentGrid);
  if(NewGrid){
    Step(dt/2, NewGrid);}
  MoveParticles(dt, CurrentGrid);
  if(NewGrid){
    Step(dt/2, NewGrid);
    Destroy(NewGrid);}}
\end{verbatim}

\noindent
Fig.~\ref{multistep} summarizes this sequence of operations, which was
proposed by Quinn \ea (1997). Whereas the coarse-grid timestep involves
accelerations calculated with all particles at the half-time point, the two
fine-grid steps involve accelerations calculated when the coarse-grid
particles are first $\Delta/4$ behind the fine-grid ones, and then ahead of
them by the same amount. The principle of the scheme is that errors arising
from these lags cancel through second order in $\Delta$.

STEP is first called on the finest domain grid with a rather large value of
$\Delta$. Through the recursive principle this call invokes calls on finer
and finer grids with smaller and smaller vales of $\Delta$ until a grid is
reached that requires no refinement, and it is advanced, so that the grid
above can be advanced, and so on. Since refinements are destroyed after
particles on them have been moved just twice, they always faithfully reflect
the particle distribution. 

The harmony of the above scheme is unfortunately marred by particles that
leave the refinement from which they started before STEP has finished. Such
departures cannot be ignored because a particle cannot continue to
contribute to the density once it is outside the grid to which it is
attached.  Consider first particles that leave their refinement at any time
up to the end of `1. fine-grid step' in Fig.~\ref{multistep}. We set the
positions and velocities of such particles back to the values they had at
$t_n$ and transfer them to the coarse grid as soon as they try to leave the
refinement (which may be at $t_{n+1/4}$ or at $t_{n+1/2}$). Hence, a
particle moves with a fine-grid time-step only if it both begins and
finishes such a fine-grid step within the refinement. Particles that leave
the refinement at $t_{n+3/4}$ during `3. fine-grid step' in
Fig.~\ref{multistep} are treated differently: such particles are immediately
transferred to the coarse grid and added to the refinement's list of
`leavers'. The forces are then evaluated at $t_{n+3/4}$, and the velocities
and positions of leavers are updated in parallel with the
coordinates of particles that remained on the refinement. Since the refinement
is destroyed at $t_{n+1}$, no significance attaches to a particle leaving
the refinement as its position is updated to $t_{n+1}$.

No special action is taken when a particle enters the space occupied by a
refinement during a call to STEP; the particle remains linked to a coarse-grid
node that has been refined, and contributes to the density on both the
coarse grid and its refinement with the spatial resolution characteristic of
the coarse grid. (For a discussion of how particles attached to a coarse grid
contribute to the density on a refinement, see Appendix B.)

The timesteps are sufficiently short that the movement of particles on grid
$n$ cannot change the density on grids $n-2$ and higher. Consequently,
drifting and kicking the particles on grid $n$ only requires mass assignment
and relaxation of the potential to be performed on grids $n-1$ and $n$, so
timesteps for the relatively small number of particles on the finest grids
are computationally inexpensive.

\subsection{Internal Units}

Let $H_0$ be the present Hubble constant, $B$ the present size of the
computational box and $\overline{\rho}$ the mean matter density. The code
uses the dimensionless variables
 \begin{equation}
 \begin{array}{lcl}
x_c&=&x/B,\\
p_c&=&p/H_0 B,\\
t_c&=&tH_0,\\
\Phi_c&=&\displaystyle\Phi H_0^2B^2,\\
\rho_c&=&\displaystyle\rho/\overline{\rho}.
 \end{array}
\end{equation} 
 In terms of these variables, the equations to be solved are
\begin{eqnarray}\label{eom}
  {\d x_c\over\d t_c}         & = &   {p_c\over a^2}, \nonumber\\
 {\d p_c\over\d t_c}         & = & -  {\nabla \Phi_c\over a}, \\
  \nabla_c \Phi_c & = &{3\Omega_{\rm M}\over2} (\rho_c - 1). \nonumber
\end{eqnarray}

\subsection{Dynamical evolution of Zel'dovich waves}\label{zeldovich}

In Section 5.2 we checked the accuracy of our Poisson solver. Here we check
our time-stepping scheme by investigating its ability to reproduce the
analytic solution for the breaking of a one-dimensional plane wave (Klypin
\& Shandarin 1983; Efstathiou \ea 1985). Since the initial conditions of a
general cosmological  simulation  are a superposition of such waves, the
ability to follow the evolution of a plane wave is a crucial test of the code.

We have used MLAPM with $\rho_{\rm ref}=1$ particle per node and Couchman's
(1991) A\pcm\ code with $\epsilon=\Delta$ with $32^3$ particles on a $32^3$
domain grid to integrate the Zel'dovich wave from the initial conditions
that are given by equation (\ref{rtoq}) and its counterpart for the momenta
 \begin{equation}\label{ZelwaveP}
p={{\dot a}^{3/2}\b k\over k^2}\cos(\b k.\b q).
\end{equation}
 Waves with three different values of $k$ were evolved with 200 timesteps on
the domain grid from $a=0.1$ until
$a=1$, when they break.  We quantify the differences between the numerical
and analytical solutions by evaluating the RMS deviations (Efstathiou~\ea
1985)
 \begin{equation}\label{errdef}
 \begin{array}{l}
  \Delta x_{\mbox{\tiny rms}} = \left[      \sum_i (x_i - x_i^{\rm a})^2 
                                      \ / \ \sum_i (x_i^{\rm a} - q_i)^2
                                \right]^{1/2} \\
\\
  \Delta v_{\mbox{\tiny rms}} = \left[      \sum_i (v_i - v_i^{\rm a})^2
                                      \ / \ \sum_i (v_i^{\rm a})^2
                                \right]^{1/2} \ ,
 \end{array}
\end{equation}
 where the super-script $a$ denotes the analytical solution
[eqs.~(\ref{rtoq}) and \ref{ZelwaveP})].  For each value of $k$
Table~\ref{testwave} shows errors from four calculations. The first and
second rows show the overall errors from A\pcm\ and MLAPM. These are broadly
comparable. The MLAPM errors contain three contributions: (i) errors in the
values of the forces at grid points; (ii) errors in the interpolation of
these forces to the locations of particles; (iii) errors in updating of
positions and momenta given the forces. The bottom row in
Table~\ref{testwave} shows that this last source of error is insignificant
by showing the errors one obtains when the force applied to each particle is
the analytic value at its location. The penultimate row shows the much
larger errors obtained when analytic forces are placed on the grid points:
evidently interpolation is a significant source of error. Since the
interpolation errors are of order a third of the overall errors shown in the
second row, there is a suggestion that we are determining the density and
then solving Poisson's equation as accurately as is profitable given the
coarseness of our grid.

\begin{table}
\caption{RMS errors in the positions ($\Delta x_{\mbox{\tiny rms}}$) and velocities
         ($\Delta v_{\mbox{\tiny rms}}$) of $32^3$ particles as defined by
		 equation (\ref{errdef})
         for AP$^3$M with $\epsilon=\Delta$ and MLAPM with a $32^3$ domain
		 grid and $\rho_{\rm
		 ref}=1$ particle per node}
\label{testwave}
 \begin{tabular}{|l||c|c|c|c|c|c|} \hline
           simulation 
                              & \multicolumn{2}{c}{$kL/2\pi=1$}
                              & \multicolumn{2}{c}{$kL/2\pi=2$}
                              & \multicolumn{2}{c}{$kL/2\pi=9$} \\ 

                              & $\Delta x$ 
                              & $\Delta v$ 
                              & $\Delta x$ 
                              & $\Delta v$ 
                              & $\Delta x$ 
                              & $\Delta v$\\ 
                              \hline \hline
         AP$^3$M              & 0.006 & 0.028 & 0.018 & 0.061 & 0.116 & 0.265 \\
         MLAPM                & 0.016 & 0.034 & 0.015 & 0.063 & 0.055 & 0.634 \\
         MLAPM(A1)            & 0.002 & 0.012 & 0.003 & 0.026 & 0.011 & 0.193 \\
         MLAPM(A2)            & 0.003 & 0.006 & 0.004 & 0.006 & 0.001 & 0.006 \\
 \end{tabular}
\end{table}

\section{\LCDM\ simulations} \label{LCDM}

In this section we explore the performance of MLAPM when used to
generate a realistic simulation.  

\begin{table}
\caption{Parameters used for three comparing simulations performed
with the ART, A\pcm, and GADGET code.\label{paramstab}}
\begin{tabular}{l|r|l|}\hline
ART & domain grid                        & $128^3$\\
    & domain steps                       & 500\\
    & $\rho_{\rm ref}$                   & 8/8\\
    & refinement level reached           & 5\\
    & number of GS sweeps on refinements & 10\\
    & CPU time                           & 47$\,$hr\\
\hline
A\pcm & softening                        & $5\hkpc$\\
      & steps                            & 4000\\
      & particles per chaining-mesh cell & 50\\
      & refinements generated            & 89\\
      & refinement level reached         & 4\\
      & CPU time                         & 69$\,$hr\\
\hline
GADGET & softening             & $5\hkpc$\\
       & velocity scale        & $10\kms$\\
       & error tolerance angle & 0.3\\
       & tree accuracy         & 0.02\\
       & tree update           & 0.05\\
       & CPU time              & 58$\,$hr\\
\hline
\end{tabular}
\end{table}

\subsection{Simulation Parameters}
We present data for simulations run with MLAPM, ART (Kravtsov \ea
1999), GADGET (Springel, Yoshida \& White 2000) and the A\pcm\ code
(Couchman 1991).  All six simulations contained $64^3$ particles
distributed through a box $15\hMpc$ on a side. The simulations started
from redshift $z=25$ with a \LCDM\ spectrum of
fluctuations. Table~\ref{paramstab} lists the parameters employed in
the ART, A\pcm\ and GADGET simulations, while Table~\ref{MLAPMtab}
gives the parameters of the three MLAPM runs. By their
end-points, all MLAPM simulations had nodes associated with grids of
$4096^3$ virtual nodes which agrees with the finest refinement level
reached in the ART run (level 5).

The parameters given in the first row of Table~\ref{MLAPMtab} are chosen to
mimic the behaviour of the ART code as closely as possible; ART is similar
to MLAPM in many ways as both codes are purely grid based. They both use a
regular domain grid covering the whole computational volume, and
sequentially refine patches of high density with finer and finer refinement
grids of arbitrary shape. The equations of motion are integrated using a
multiple time stepping scheme that employs half the time step of the
previous level on every given refinement.  But there are subtle differences,
too. The first, most obvious difference is the way the solution is obtained
on the finest domain grid: ART uses an FFT solver whereas MLAPM utilizes
Brandt's multigrid scheme (Brandt 1977). Moreover, MLAPM uses the
Triangular-Shape-Cloud (TSC) mass assignment scheme in contrast to the
Cloud-In-Cell (CIC) scheme applied by ART. The equations of motion in the
ART code are integrated using the expansion factor $a$ as integration
variable, which was also applied to MLAPM's `run~$a$' to make those two runs
as similar as possible.  Two other MLAPM runs ($t$1 and $t$2) use time~$t$
for integrating the equations of motion (cf. Eq.~\ref{spHamil} in
Section~\ref{SECstepping}) and perform 50 percent more Gauss-Seidel sweeps
on each grid before checking for convergence.  The latter results in a lower
performance in terms of time, but should lead to more accurate solutions of
Poisson's equation. We will investigate these propositions in more detail
below.

\begin{figure*}
\centerline{\psfig{file=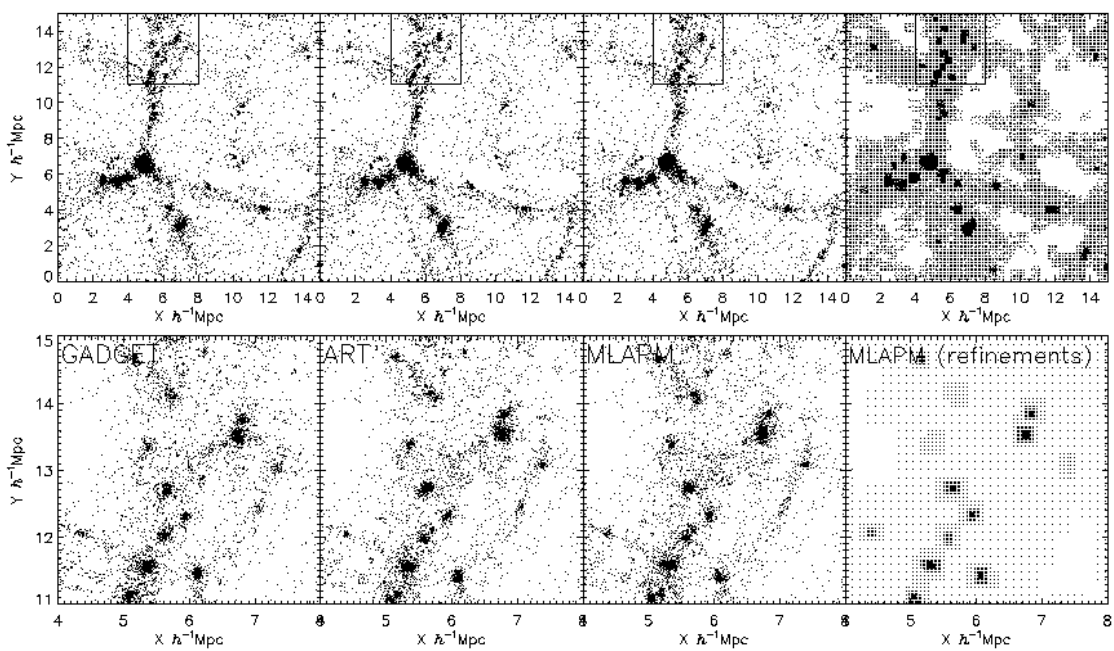,width=\hsize}}
\caption{Slice through three \LCDM\ simulations run from identical
initial conditions using (from left to right) GADGET, ART, and MLAPM
(run~$a$). The extreme right-hand panels show the final grid hierarchy
of the MLAPM simulation. The bottom panels show enlargements of the
small region marked in the upper panels.\label{slicefig}}
\end{figure*}

\begin{table}
\caption{Parameters of MLAPM simulations. The value for the number of
domain grid cells is given in the second column and the number of
integration steps also applies for the domain grid. The first number
in the $\rho_{\rm ref}$  column is the refinement density on the domain
grid, while the second number applies to all finer grids. The same
convention applies in the column headed `GS sweeps'.  The
simulation plotted in Fig.~\ref{slicefig} corresponds to run~$a$ which 
parameters were chosen as close as possible to the ones in the ART run.
\label{MLAPMtab}}
\begin{tabular}{c|c|c|c|c|c|}\hline
run  &    grid  & $\rho_{\rm ref}$ &   steps   & GS sweeps & CPU time\\ \hline
 $a$  &    $128^3$  &          8/8     &     500      &   10/10   &  $42\,$hr\\
 $t$1 &    $128^3$  &          8/8     &     500      &   15/15   &  $69\,$hr\\
 $t$2 &    $64^3$   &          1/8     &     250      &   15/15   &  $48\,$hr\\
\end{tabular}
\end{table}

\subsection{Comparisons}
Fig.~\ref{slicefig} shows slices through the GADGET and ART
simulations, and the MLAPM simulation that corresponds to the first
row of Table~\ref{MLAPMtab} (run~$a$).  The lower panels show
enlargements of a small region of the upper panels. The rightmost
panels show the final  grid
structure of the MLAPM simulation.  The three particle distributions
are clearly very similar, but not identical. In comparisons between
simulations run with A\pcm\ and ART, Knebe et al.\ (2000) detected
similar differences, and showed that understanding the physical
significance of these differences is not straightforward. In
particular, simulations run with different codes tend to be at
slightly different phases at a given time. Such phase differences are
probably not physically significant, but can lead to material
differences in the appearance of slices such as those shown in
Fig.~\ref{slicefig}. For example, in one panel a small cluster may be
evident while in another it is invisible because its centre lies just
above or below the slice shown.

\begin{figure}
\centerline{\psfig{file=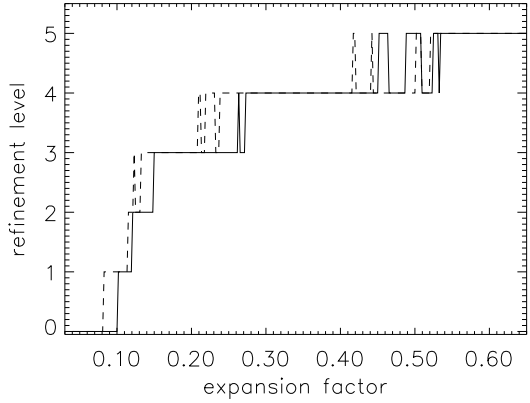,width=\hsize}}
\caption{Refinement levels invoked by MLAPM (run~$a$) and ART (dashed curve) as 
         a function of the expansion factor.\label{levelfig}}
\end{figure}

It is now interesting to compare MLAPM (run~$a$) with the ART run as both
are set up as similarly as possible. In Fig.~\ref{levelfig} we therefore
plot for both codes the refinement level reached against the expansion
factor $a$. From $a\sim 0.55$ onwards no finer refinements are generated and
hence there is no need to extend the plotted data to $a=1$. We can clearly
see that both codes start using the same refinements at about the same time,
with ART creating its levels slightly earlier. Otherwise the curves agree
fairly well, demonstrating how similarly MLAPM and ART are dealing with
refinements. The `noisy behaviour' can be ascribed to the small size of
refinements when they are first created; all adaptive grids are placed
around initially small high density regions, which might fluctuate around
the density threshold for a couple of steps until stabilized. To compensate
for this effect MLAPM refines at the beginning of each domain grid step down
to the actually needed refinement level but does not allow finer grids to be
called into existence during the course of that  domain step. This
might explain why MLAPM's refinements appear to be invoked slightly later
than ART's (Fig.~\ref{levelfig}).

\begin{figure}
\centerline{\psfig{file=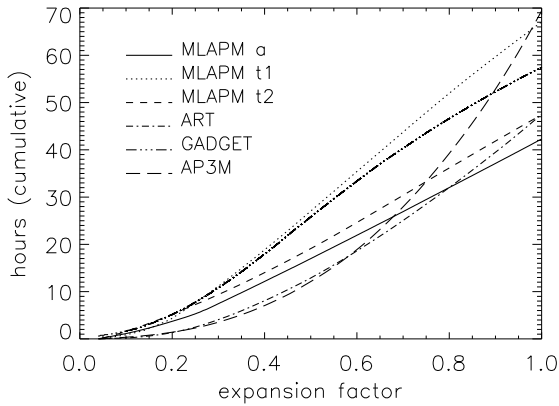,width=\hsize}}
\caption{CPU time used as a function of the expansion factor reached for the
simulations shown in Fig.~\ref{slicefig} and three other MLAPM simulations.
The curve for the MLAPM simulation plotted in Fig.~\ref{slicefig} finishes
second from the bottom.\label{timefig}}
\end{figure}

Fig.~\ref{timefig} shows, again as a function of expansion
parameter~$a$, the CPU time required by all six simulations. Since the
speed with which a given code runs depends sensitively on the values
chosen for its various (technical) parameters, exact comparisons are
difficult to make.  Experiments with slightly modified parameters for
ART, GADGET, and A\pcm\ showed that the total times needed to run a
simulation can vary by up to 50\%  without perceptible
change in the statistical analysis as given below. The only
difference between MLAPM's run~$a$ and run~$t$1, besides the
integration variable, is the number of GS sweeps performed on each grid
before checking for convergence.  As most of the time is spent on
solving Poisson's equation, we get an increase of more than 60\% in
time when using 50\% more sweeps; we also observe slightly bigger
refinements in run~$t$1 which accounts for the remaining 10\% decrease
in performance.

It is also worth noticing that A\pcm\ and ART both perform similarly
at early times, when the forces are (mainly) based on an FFT solver. Only
when particles start to cluster and the PP part becomes more and more
important in the A\pcm\ run does ART start to show its advantage by using
arbitrarily shaped refinements in high density regions to increase the force
resolution. However, MLAPM overtakes ART at times, when the use of
refinements is dominating the time budget. This behaviour suggests that our
(de-)refinement procedure is more time efficient and indicates that the
difference in performance at early times between ART and MLAPM can be
ascribed to our adoption of Brandt's multigrid plan even on the domain
grid. But again, checking the relative timings of an FFT solver and the
multi-level algorithm is a job for the future.

\begin{figure}
\centerline{\psfig{file=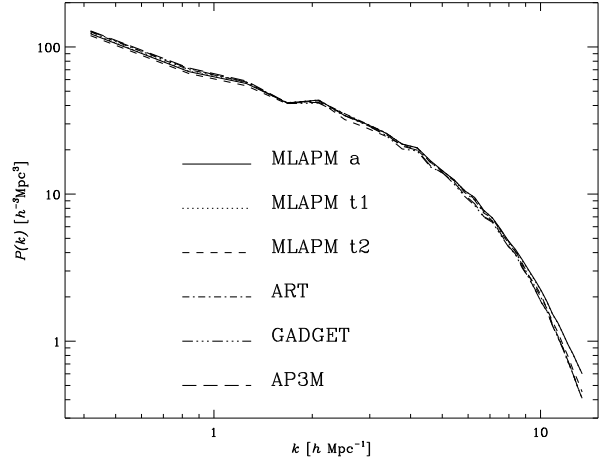,width=\hsize}}
\caption{Power spectra at redshift $z=0$ for all six simulations.
The lower solid line is actually a superposition of broken lines.
\label{powerfig}}
\end{figure}

In Fig.~\ref{powerfig} we show the dark matter power spectra of all
six simulations at a redshift $z=0$. There are no obvious differences
and they all agree very well with each other.

\begin{figure}
\centerline{\psfig{file=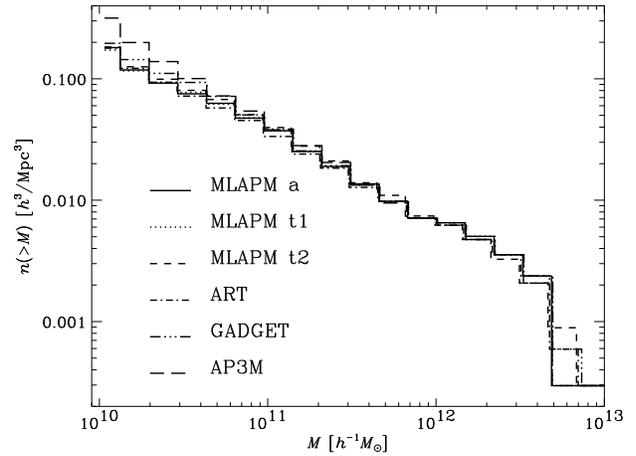,width=\hsize}}
\caption{Mass functions at redshift $z=0$ for all six simulations.
         Halos were identified using a standard friends-of-friends
         algorithm.
\label{massfig}}
\end{figure}

However, when investigating the cumulative mass function $n(>M)$ for
particle groups (Fig.~\ref{massfig}) identified using a standard
friends-of-friends group finder with linking length 0.17 (which
corresponds to an overdensity of about 330), there are subtle
deviations between the runs. At the high mass end they all coincide,
but at the low mass end of the distribution function we observe more
small objects in A\pcm\ and (to a smaller extent) GADGET than in any of the other runs. This
agrees with findings by Knebe \ea (2000), where it was shown that
A\pcm\ tends to form more low-mass objects in underdense regions
(cf. Fig.~3 in that paper).

\begin{figure}
\centerline{\psfig{file=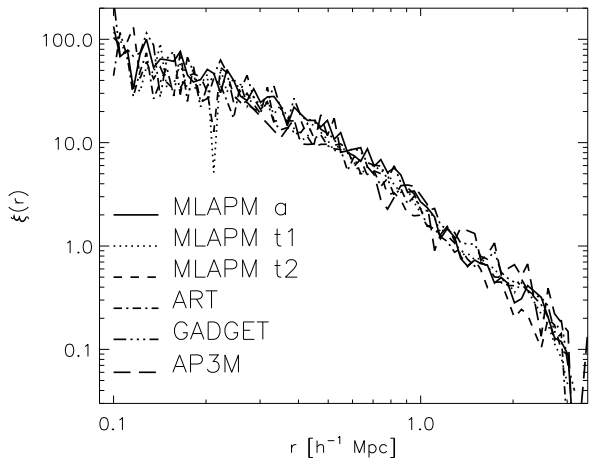,width=\hsize}}
\caption{Halo-Halo correlation functions at redshift $z=0$ for all six
         simulations.  Halos were identified using a standard
         friends-of-friends algorithm. 
\label{xifig}}
\end{figure}

Finally we show the halo-halo correlation function for the
objects presented in Fig.~\ref{massfig} -- the agreement between the
codes is good.

These comparisons convince us that all four codes produce comparable results
in comparable times, except that there are small differences in the mass
functions produced by grid-based methods (MLAPM and ART) and PP-based ones
(A\pcm\ and GADGET).  We also find that there are only modest changes in the
scientific results when fiddling with the technical parameters, i.e. the
number of GS sweeps.

\subsection{MLAPM performance}
 This section deals with the dependence of MLAPM's performance on the values
taken by technical parameters and the way the grids are used. 

\begin{figure}
\centerline{\psfig{file=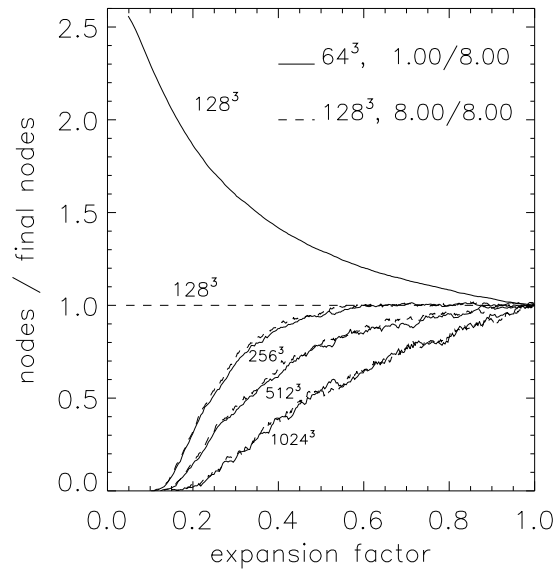,width=\hsize}}
\caption{Numbers of nodes at each level of refinement as a function of
expansion factor. Grids as fine as $4096^3$ virtual nodes are created, but
data for the finest two grids are not plotted.
\label{nodefig}}
\end{figure}

Fig.~\ref{nodefig} shows as a function of expansion factor achieved
the numbers of nodes at each refinement level for two simulations:
those listed in the second and third rows of Table~\ref{MLAPMtab}
(runs~$t$1 and~$t$2). The growth in the grids with $256^3$ or more
virtual nodes is identical in the two simulations and the last two
levels ($2048^3$ and $4096^3$) are not shown for clarity. When the
domain grid has $64^3$ nodes, the number of nodes in the $128^3$ grid
falls by a factor 2.5 during the simulation, as particles drain out of
voids and more and more domain-grid nodes fail to achieve the
threshold for refinement, namely $\rho_{\rm ref}=1.0$ particles per
node.

Comparison of the timings listed in the lower two rows of
Table~\ref{MLAPMtab} (run~$t$1 and $t$2) shows that MLAPM is slowed
when the number of domain grid cells is changed from $64^3$ to
$128^3$. This is explained in Table~\ref{steptab}, where we show the
average time spent on a given grid over the course of the whole
simulation. Note, that in run~$t$2 we are also doing 500 steps on the
$128^3$ grid and our refinement criterion was chosen to 
refine (nearly) the whole domain grid at early times.  As this run requires
both less CPU time and less memory, it provides better value for
money.

\begin{table}
\caption{Average CPU time in seconds per step over the course of a
simulation with $64^3$ particles on a $64^3$ domain grid (run~$t$2) and
$128^3$ domain grid (run~$t$1), respectively.\label{steptab}}
\begin{tabular}{l|r|r|r|r|r|}
Grid:           & 64   & 128  & 256 & 512 & 1024\\ \hline
MLAPM $t$1:       & ---  & 261  & 18  & 5   & 2\\
MLAPM $t$2:       & 35   & 127  & 21  & 5   & 2\\
\end{tabular}
\end{table}

It is interesting to see how the CPU time required per step varies
between grids. Again, Table \ref{steptab} lists the average CPU time
per step used to solve for the forces and move the particles on the
first through fourth refinements in the course of the simulation whose
end-point is plotted in the rightmost four panels of
Fig.~\ref{slicefig}. It is always the $128^3$ grid which dominates the
time budget, but even when we add up the time spent on the $64^3$ and
the $128^3$ grid for run~$t$2 we are still faster than using a regular
$128^3$ grid all the time, because at later times there are far
fewer nodes to sweep over (cf. Fig.~\ref{nodefig}).  Thus the enhanced
resolution that an adaptive grid provides in high density regions
comes at an insignificant cost in both memory and CPU time.

\subsection{Layzer--Irvine equation}

A useful check on the accuracy of a cosmological simulation is provided by
the Layzer--Irvine equation. To derive an appropriate form of this, we
assume that
the single-particle potential $\Phi$ that appears in the
Hamiltonian (\ref{spHamil}) could be obtained
by a sum over pairs of some time-independent smoothing kernel $S$. With this
assumption  the total potential energy of the system is
 \begin{eqnarray}
{\displaystyle{U\over a}}&=&{\displaystyle
{1\over2a}\sum_{\alpha=1}^N\Phi(\b x_\alpha)}\nonumber\\
&=&{\displaystyle
-{1\over2a}\sum_{\alpha\neq\beta}S(|\b x_\alpha-\b
x_\beta|),}
\end{eqnarray}
 where the sum is over all particles and the coordinates are comoving ones.
It is straightforward to check that our equations of motion (\ref{He}) can
be obtained from the $N$-body Hamiltonian
\begin{equation}
H={K\over a^2}+{U\over a},
\end{equation}
 where
\begin{equation}
K\equiv\fracj12\sum_{\alpha=1}^N p_\alpha^2.
\end{equation}
 We have
\begin{eqnarray}\label{dHdt}
{\d H\over\d t}&=&{\pa H\over\pa t}\nonumber\\
&=&-{\dot a\over a}\left({2K\over a^2}+{U\over a}\right).
\end{eqnarray}
 Consequently
 \begin{equation}
\left[H\right]_{t_1}^{t_2}=-\int_{t_1}^{t_2}{\d a\over a}\left({2K\over
a^2}+{U\over a}\right).
\end{equation}
 This equation, which is valid no matter how $a$ depends on time, states
that the Hamiltonian would be constant if the system were fully virialized.
For some reason it is conventional not to monitor the satisfaction of this
equation, but of an alternative conservation equation that follows from
equation (\ref{dHdt}), namely
\begin{eqnarray}
{\d aH\over\d t}&=&\dot aH+a\dot H\nonumber\\
&=&-\dot a{K\over a^2}\ .
\end{eqnarray}
 Consequently 
\begin{equation}\label{LIeq}
C\equiv [aH]_{t_1}^{t_2}+\int_{a_1}^{a_2}\d a\,{K\over a^2}
\end{equation}
 should be constant. 

\begin{figure}
\centerline{\psfig{file=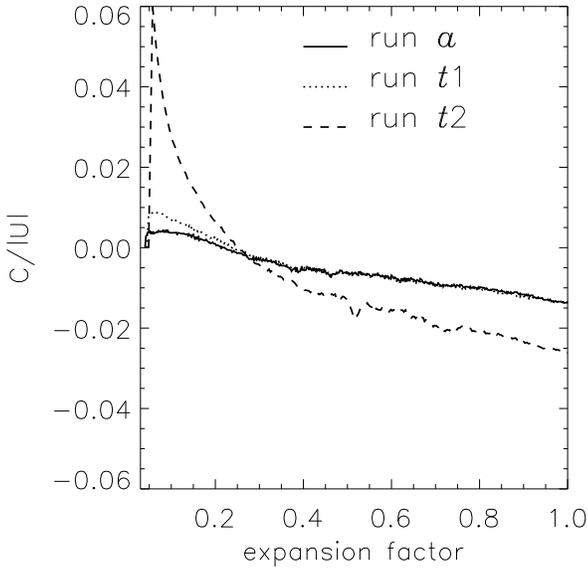,width=\hsize}}
\caption{Variation with $a$ of the  Layzer-Irvine invariant $C$ that is defined by
equation (\ref{LIeq}).\label{LIfig}}
\end{figure}

Fig.~\ref{LIfig} plots $C/|U|$ as a function of $a$ for all three MLAPM
simulations. By taking smaller time-steps one can show that truncation error
in the integration of the equations of motion (\ref{He}) makes a negligible
contribution to the variation of $C$.  Errors in interpolating the forces
from the grid to the locations of particles (see Section 5.1.3) cause $C$ to
vary by causing the force on a particle to differ slightly from the local
potential gradient.  Another important contributor to the variability of $C$
is the fact that $\Phi$ cannot be obtained from a time-independent smoothing
kernel $S(|\b x|)$, as we assumed in deriving equation (\ref{LIeq}). Indeed,
no such kernel would give our potential precisely, because our softening
length $\epsilon$ diminishes from voids to the cores of clusters.  Moreover,
the mean value of $\epsilon$ diminishes over time as clustering develops and
finer and finer grids are created.  Since the Layzer--Irvine equation is
based on the assumption of a spatially and temporally invariant softening
kernel, it follows that variability of $\epsilon$, for which
Section~\ref{SECperformance} presents a powerful case, will lead to
significant violations of the Layzer--Irvine equation. This is also
reflected in the curve for run~$t$2 as in this case we started out with a
$64^3$ domain grid and used a grid with $128^3$ virtual nodes that varied
significantly in extent (cf. Fig.~\ref{nodefig}). However, the variation in
$C/|U|$ is less than 2 to 3 percent except for a steep
rise in $C$ during the very first steps, which is probably due to sharp
changes in the resolution provided by the  $128^3$ grid. As the calculation
settles down there follows  a
more moderate decrease in $C/|U|$, which finishes at  about 2 percent.

\section{Discussion and Conclusions}\label{discussion}

In the coming years work with cosmological simulations will increasingly
focus on the formation of galaxies of various types. Such simulations demand
the highest attainable spatial and mass resolution and will stretch
available computer power to its limits. The efficiency of the available
computer codes, both in respect of CPU time and memory usage, will be of
paramount importance. $N$-body codes can be divided into those that find the
gravitational potential by summation over a Green's function, and those that
solve Poisson's equation on a grid. To be efficient, the grid employed in
the latter type of code has to be capable of adapting itself dynamically to
the evolving mass distribution, and this requirement leaves one little
option but to solve Poisson's equation by Brandt's multigrid technique.

The code presented here, MLAPM, is one of two cosmological codes that
deploys such a grid, the other being the ART code (Kravtsov et al.~1997, 1999).
Both codes subdivide cells in which the density exceeds a threshold, and
move particles with timesteps that decrease by a factor 2 with each
additional level of refinement of the region within which they lie. With these
codes gravity is automatically softened adaptively, so that the softening
length is near its optimum value in both high and low-density regions. With
A\pcm\ and most tree codes, by contrast, a single softening length is
employed at all times and places, with the result that it is generally much
smaller than it should be in low-density regions.

Although MLAPM and ART are conceptually very similar, they do differ in a
number of important respects. In particular,

\begin{itemize} 
\item MLAPM uses a simple recursive and fully symplectic integration scheme;

\item since MLAPM is written in $C$
rather than {\tt FORTRAN}, it can make extensive use of dynamic memory
allocation;

 \item MLAPM uses Brandt's multi-grid approach for solving
       Poisson's equation even on the domain grid, whereas
       the ART code uses an FFT solver.

\end{itemize}

MLAPM has a single free parameter, the threshold density for node
refinement, $\rho_{\rm ref}$. Smaller values of $\rho_{\rm ref}$ yield finer
grids and harder forces. The memory used by grids is proportional to
$\rho_{\rm ref}^{-1}$ and exceeds the memory used by particles for
$\rho_{\rm ref}\lta8$ particles per node. 

Tests of the ability of the code to recover the gravitational fields of
virialized structures and strongly non-linear plane waves show that
radically different values of $\rho_{\rm ref}$ are required in the two
cases. With $\rho_{\rm ref}<8$ particles per node, forces near the centre of
a typical virialized structure fluctuate from realization to realization by
more than 25 percent. Hence, 8 particles per node seems a minimum value for
$\rho_{\rm ref}$ when representing virialized halos.

By contrast, to recover a reasonable approximation to the field of a wave
whose frequency exceeds half the Nyquist frequency of the domain grid with
as many particles as the grid has nodes, we require $\rho_{\rm ref}\lta1$
particle per node, which ensures that the domain grid is refined through
most of the simulation. Such a small value works well prior to virialization
although it would yield a completely noise-dominated gravitational field
after it, because the usual procedure for setting up the initial conditions
of cosmological simulations enables the underlying density to be recovered
from the particle positions, free of Poisson noise. 

In view of the different refinement criteria required before and after
virialization, one of two strategies should be adopted. In the first one
uses a domain grid with as many nodes as there are particles, and on it sets
$\rho_{\rm ref}$ to a value less than unity. This ensures that the domain
grid is refined everywhere until voids develop in which the particle density
is low enough for adequate resolution to be provided by the domain grid
alone.  In the second strategy, the domain grid has eight times as many
nodes as there are particles, and one sets $\rho_{\rm ref}=8$ particles per
node on every grid. The second strategy is safer unless
clustering is so highly developed that the density is less than an eighth of
the mean density in a significant volume. However, our experiments show that
using a coarse domain grid with $\rho_{\rm ref}=1$ yields statistically
indistinguishable results at lower cost than the conservative strategy.

Before a code for cosmological simulations can now be considered
complete, it should include instructions written in MPI that will
enable it to run on a distributed-memory multi-processor computer. To
our knowledge only one such code is currently publicly available for
cosmological $N$-body simulations, the tree code GADGET (Springel, \ea
2000). Producing an MPI version of MLAPM is a high
priority. Multigrid codes are in principle well suited to
parallelization because each subgrid of the domain grid can be
advanced substantially independently of the others. Moreover, the data
structures (nodes and quads) associated with physically connected
nodes are already allocated in a way that makes them likely to be
stored in adjacent blocks of memory, and the existing linking of
particles to nodes means that it would be simple to ensure that data
for physically connected particles were always stored together. The
only significant problem we anticipate encountering in the
parallelization arises from the recursive nature of the calls to
step. Such recursive calls are known to be a barrier to
parallelization. Fortunately, by making a few copies of STEP, called
STEP0, STEP1, \dots, or whatever, it is trivial (if inelegant) to make
the algorithm non-recursive, at least for the first few calls. Loops
over $z$QUADS within one of these non-recursive copies of STEP could
then be parallelized.

\section*{Acknowledgments}
 AK thanks Stefan Gottl\"ober for many useful and encouraging
discussion, and thanks Anatoly Klypin and Andrey Kravtsov for valuable
comments and kindly providing a copy of the ART code.  We are grateful to
Hugh Couchman for permitting us to use the A\pcm\ code and to Volker
Springel for access to the GADGET code. JJB thanks the Astronomy Department
of the University of Washington for hospitality during the drafting of this
paper.  He was then supported in part by NSF grant AST-9979891. This work
has benefited from the facilities of the Oxford Supercomputer Centre.


\section*{Appendix A: Density from mass on a Zel'dovich-distorted grid}

We show that a distribution of particles placed on a Zel'dovich distorted
grid uniquely defines the underlying density field.
Consider the generalization of equation (\ref{rtoq}) to many
waves, one for each site of a lattice in $\b k$-space. We have
that  the $N$ particle displacements $\b e_\alpha=\b
r_\alpha-\b q_\alpha$ are related to the amplitudes $A_\b k$ of the
generating waves by the finite sum
 \begin{equation}
\b e_\alpha=\sum_{|\b k|<K}\b A_\b k\cos(\b k.\b q_\alpha).
\end{equation}
 If the particles are on the grid in $\b q$-space that is the reciprocal of
the $\b k$-space grid, then this equation states that the $\b e_k$ are
related to the $\b A_k$ by a DFT.  Hence we can recover the latter from the
particle positions and then
reconstruct the entire density field from the Jacobian
 \begin{equation}\label{Jaceq}
\rho=\overline{\rho}\left[{\pa(\b r)\over\pa(\b q)}\right]^{-1}.
\end{equation}

We have investigated the possibility of obtaining the density in regions
that have yet to virialize from equation (\ref{Jaceq}). We find that
numerical differentiation of $\b r$ with respect to $\b q$ does yield more
accurate values of the density than the TSC mass-assignment scheme,
especially in voids. However, most of this advantage is lost in propagating
the density from the locations of particles to nodes.

\section*{Appendix B: Particle transfer to fine grids}

\begin{figure}
\centerline{\psfig{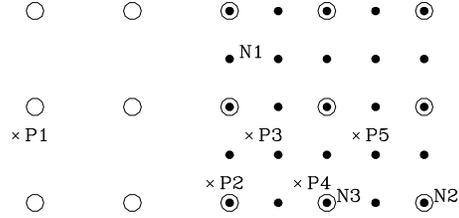}}
\caption{Particles and nodes at the edge of a refinement.\label{transfig}}
\end{figure}

We describe intricacies that arise when particles transfer from
coarser to finer grids. Fig.~\ref{transfig} shows five particles at or
near the boundary of a refinement (which always includes fine nodes
that are cospatial with coarse nodes). Assigning the masses of a
particle such as P1 is straightforward because its mass only
contributes to the density on the coarse grid. Similarly, the mass of
P5 only contributes to the density on the fine grid. The masses of the
other three particles contribute to the density on both grids, and
considerable care has to be exercised in its assignment.  The main
problem is to determine the contributions to the fine grid of
particles like P2 and P3 that remain on the coarse grid.

We start by transferring particles P4 and P5 to the fine grid. Next we
use the coarse grid's TSC kernel to subtract from coarse-grid nodes
the mass of these particles, and the fine grid's kernel to add the
same mass to fine-grid nodes. At this stage mass is assigned to
boundary nodes such as N1. When this has been done the density on a
coarse-grid node such as N2 that lies in the interior of the
refinement will be zero, while a coarse-grid node such as N3 that lies
near the refinement's edge will have non-zero density. In fact, the
density on N3 will come from particles such as P2 and P3. We use the
fine grid's TSC kernel to distribute this mass among the neighbouring
fine-grid nodes, but for the present we hold it in temporary
variables, separate from the masses associated with P4 and P5. Now we
use the restriction operator to add the fine-grid density to all
coarse-grid nodes that are cospatial with a fine-grid node. This
operation completes the determination of the coarse-grid
density. Finally we complete the determination of the fine-grid
density by adding to each fine-grid node the mass held in its
temporary variable.

\end{document}